\documentclass[useAMS,usenatbib]{style/mnras}
\usepackage{epsfig,amsmath,natbib,wasysym}
\usepackage{color}
  
\def\refjnl#1{{\rm#1}}
\def\apjl{\refjnl{ApJ}}                
\def\apjs{\refjnl{ApJS}}               
\def\ao{\refjnl{Appl.~Opt.}}           
\def\aap{\refjnl{A\&A}}                

\def\refjnl#1{{\rm#1}}
\def\apjl{\refjnl{ApJ}}                
\def\apjs{\refjnl{ApJS}}               
\def\ao{\refjnl{Appl.~Opt.}}           
\def\aap{\refjnl{A\&A}}                

\def\be{\begin{equation}} 
\def\ee{\end{equation}} 
\def\ba{\begin{eqnarray}} 
\def\ea{\end{eqnarray}}

\def\cc{\,{\rm {cm^{-3}}}} 
\def\msun{{\Msun}}

\def\gsim{\lower.5ex\hbox{\gtsima}} 
\def\lsim{\lower.5ex\hbox{\ltsima}} \def\gtsima{$\; \buildrel > \over 
\sim \;$} \def\ltsima{$\; \buildrel < \over \sim \;$} \def\prosima{$\; 
\buildrel \propto \over \sim \;$} \def\gsim{\lower.5ex\hbox{\gtsima}} 
\def\lsim{\lower.5ex\hbox{\ltsima}} 
\def\simgt{\lower.5ex\hbox{\gtsima}} 
\def\simlt{\lower.5ex\hbox{\ltsima}} 
\def\simpr{\lower.5ex\hbox{\prosima}} \def\la{\lsim} \def\ga{\gsim} 
  
 \def\gtsima{$\; \buildrel > \over \sim \;$} 
\def\ltsima{$\; \buildrel < \over \sim \;$} 
\def\gsim{\lower.5ex\hbox{\gtsima}} 
\def\lsim{\lower.5ex\hbox{\ltsima}} 
\def\simgt{\lower.5ex\hbox{\gtsima}} 
\def\simlt{\lower.5ex\hbox{\ltsima}} 
\def\simpr{\lower.5ex\hbox{\prosima}} 
\def\la{\lsim} 
\def\ga{\gsim}

\def\msun{\,{\rm \Msun}}

\def\E3{{\cal E}_{\rm g}^{III}}

\def\r12{r_{1/2}} 
\def\x12{x_{1/2}} 
\def\v12{v_{1/2}}

\usepackage{color}
\usepackage{amssymb}
\usepackage{multirow}

\newcommand{\quotes}[1]{``#1''}

\hypersetup{draft=false}

\defcitealias{Gorti2002}{G02}
\defcitealias{Cicone2014}{C14}

\def\G02cit{\citetalias{Gorti2002}}
\def\C14cit{\citetalias{Cicone2014}}

\def\lsun{{\rm L}_{\odot}}
\def\msun{{\rm M}_{\odot}}
\def\rsun{{\rm R}_{\odot}}
\def\h2{{\textsc{h}_2}}
\def\hii{\textsc{hii}}
\def\hi{\textsc{hi}}
\def\icm{\textsc{icm}}
\def\pdr{\textsc{pdr}}

\title[Molecular clumps photoevaporation in ionized regions]{Molecular clumps photoevaporation in ionized regions} 
\author[Decataldo et al.]{D. Decataldo$^{1}$, A. Ferrara$^{1,2}$, A. Pallottini$^{3,1,4,5}$, S. Gallerani$^{1}$, L. Vallini$^{6}$ \\
$^{1}$ Scuola Normale Superiore, Piazza dei Cavalieri 7, I-56126 Pisa, Italy\\
$^{2}$ Kavli IPMU, The University of Tokyo, 5-1-5 Kashiwanoha, Kashiwa 277-8583, Japan\\
$^{3}$ Centro Fermi, Museo Storico della Fisica e Centro Studi e Ricerche ``Enrico Fermi'', Piazza del Viminale 1, Roma, 00184, Italy\\
$^{4}$ Kavli Institute for Cosmology, University of Cambridge, Madingley Road, Cambridge CB3 0HA, UK\\
$^{5}$ Cavendish Laboratory, University of Cambridge, 19 J. J. Thomson Ave., Cambridge CB3 0HE, UK\\
$^{6}$ Nordita, KTH Royal Institute of Technology and Stockholm University, Roslagstullsbacken 23, SE-10691 Stockholm, Sweden}

\begin{document}
 
\date{\today} 
 
\pagerange{\pageref{firstpage}--\pageref{lastpage}} \pubyear{2017}
 
\maketitle 

\setlength{\parskip}{0pt}
    
\label{firstpage} 
\begin{abstract} 
We study the photoevaporation of molecular clumps exposed to a UV radiation field including hydrogen-ionizing photons ($h\nu > 13.6$ eV) produced by massive stars or quasars. We follow the propagation and collision of shock waves inside clumps and take into account self-shielding effects, determining the evolution of clump size and density with time. The structure of the ionization-photodissociation region (iPDR) is obtained for different initial clump masses ($M=0.01 - 10^4\,\msun$) and impinging fluxes ($G_0=10^2 - 10^5$ in units of the Habing flux). The cases of molecular clumps engulfed in the HII region of an OB star and clumps carried within quasar outflows are treated separately. We find that the clump undergoes in both cases an initial shock-contraction phase and a following expansion phase, which lets the radiation penetrate in until the clump is completely evaporated. Typical evaporation time-scales are $\simeq 0.01$ Myr in the stellar case and 0.1 Myr in the quasar case, where the clump mass is 0.1 $\msun$ and $10^3\,\msun$ respectively. We find that clump lifetimes in quasar outflows are compatible with their observed extension, suggesting that photoevaporation is the main mechanism regulating the size of molecular outflows.
\end{abstract}

\begin{keywords}
ISM: clouds, evolution, photodissociation region - quasars: general
\end{keywords}

\section{Introduction}
\label{Intro}

The diffuse interstellar medium (ISM) is characterized by a turbulent multi-phase structure, showing a broad range of densities, temperatures and chemical compositions. In some regions, gravitational forces and pressure compress the gas to sufficiently high densities, so that the formation of molecules such as H$_2$ and CO is allowed.     

CO maps have revealed that Giant Molecular Clouds (GMCs) contain a very rich internal structure featuring filaments and clumps \citep{Bally1987, Bertoldi1992}. The typical sizes of the detected clumps range from 1 to 10 pc. Temperature and density of the gas can be estimated by combining line intensities with radiative transfer calculations. Such studies yield kinetic temperatures in the range $T=15-200$ K, associated with H$_2$ densities of $n=10^{3-4}$ cm$^{-3}$ \citep{Parsons2012, Minamidani2011}. A correlation between clump temperature and H$\alpha$ flux suggests that denser clumps are warmer because of a larger UV radiation intensity, likely provided by external sources or internal star-formation activity.

Dense molecular clumps have also been detected within the Photo Dissociation Regions (PDRs) of OB stars, through observations in the infrared and millimiter bands \citep{VanderWerf1993, Luhman1998}. Detections of fine-structure lines of [CI] and [CII], high-$J$ CO rotational lines, and $J=3-2$ lines of HCN and HCO, show that PDRs are made of a low-density, more diffuse component ($n\simeq 10^{2-4} $ cm$^{-3}$), and high-density structures ($n\simeq 10^{6-7}$ cm$^{-3}$), such as in M17SW \citep{Hobson1992, Howe2000}, and in the Orion bar \citep{Lis2003}. These clumps must have sizes as small as one tenth or a hundredth of pc, often showing elongated shapes. The presence of such clumps affects significantly the emission spectrum of stellar PDRs.

According to recent observations \citep{Cicone2014}, molecular clumps are also detected in outflowing gas around quasars. The radiation pressure drives a powerful wind ($v\sim 0.1c$ with $c$ speed of light) which collides with the ISM, so that a shock propagates forward into the ISM and a reverse shock propagates back into the wind \citep[model by][]{King2010}. The outflowing gas is heated by the shock to very high temperatures ($T\sim 10^7$ K), so that it is expected to be completely ionized. Nevertheless, detections of the CO, OH and H$_2$O lines \citep[e.g.][]{Alatalo2011, Aalto2012, Feruglio2015} show that the outflow is in molecular form up to a radius of $1-10$ kpc. To reach such distances, the molecular gas has to be structured in clumps, able to provide sufficient self-shielding against the strong quasar radiation field. 

The structure of a molecular clump is significantly determined by the presence of an ionizing/photo-dissociating radiation field, since incident photons with different wavelengths alter the chemical composition of the gas and its physical properties. Far ultraviolet (FUV) radiation ($6$ eV$< h \nu< 13.6$ eV) is responsible for the dissociation of molecules, determining the formation of a PDR \citep{Tielens1985, Kaufman1999, Rollig2007, Bisbas2012} at the surface of the clump itself. Furthermore, radiation above the Lyman limit ($h\nu > 13.6$\, eV) ionizes neutral atoms, and it is completely absorbed within a shallow layer. 

The goal of this paper is to understand the evolution of radius and the density profile of molecular clumps exposed to a UV radiation field including hydrogen-ionizing photons produced by massive stars or quasars. The key point is that an ionized shell \textit{and} an atomic shell form at the edge of the clump. The dynamics of this layered structure is determined by the fact that each layer is at a different temperature and pressure. For a clump with initial density $n_0\simeq10^5$ cm$^{-3}$, typical temperatures deep into the clump are $T\simeq 10-100$ K, while an atomic (ionized) region can be heated up to around $T\simeq 10^3$ K ($T\simeq 10^4$ K). We denote this type of ionization/photodissociation regions as iPDR.

In particular, we apply our model to two scenarios. 
\begin{itemize}
	\item Stellar case: a molecular clump is in pressure equilibrium within a GMC in the proximity of an OB star, and it is suddenly engulfed by the expanding HII region.
	\item Quasar case: a clump forms as a result of thermal instabilities within the outflow, finds itself embedded in the ionized wind and exposed to the quasar radiation.  
\end{itemize}
Previous works in the literature concentrated mostly on the effects of non-ionizing photons on photoevaporation of clouds \citep{Gorti2002, Adams2004, Champion2017}. The evolution in their case is simplified by the fact that the clump is exposed only to radiation below the Lyman limit. As a result, the clump develops a single shell structure.

The paper is organized as follows. In Sec. \ref{Met} we describe the model adopted for the structure of molecular clumps, together with the physics of the dynamical and thermal processes involved. In Sec. \ref{Res} we present the results for the evolution of the clump radius, and compute the evaporation time of a clump for different parameters of the system. The results are presented separately for clumps located near stellar or quasar sources. In Sec. \ref{Con} we summarize our results. 

\section{Model}
\label{Met}

\begin{figure}
\centering
\includegraphics[width=0.45\textwidth]{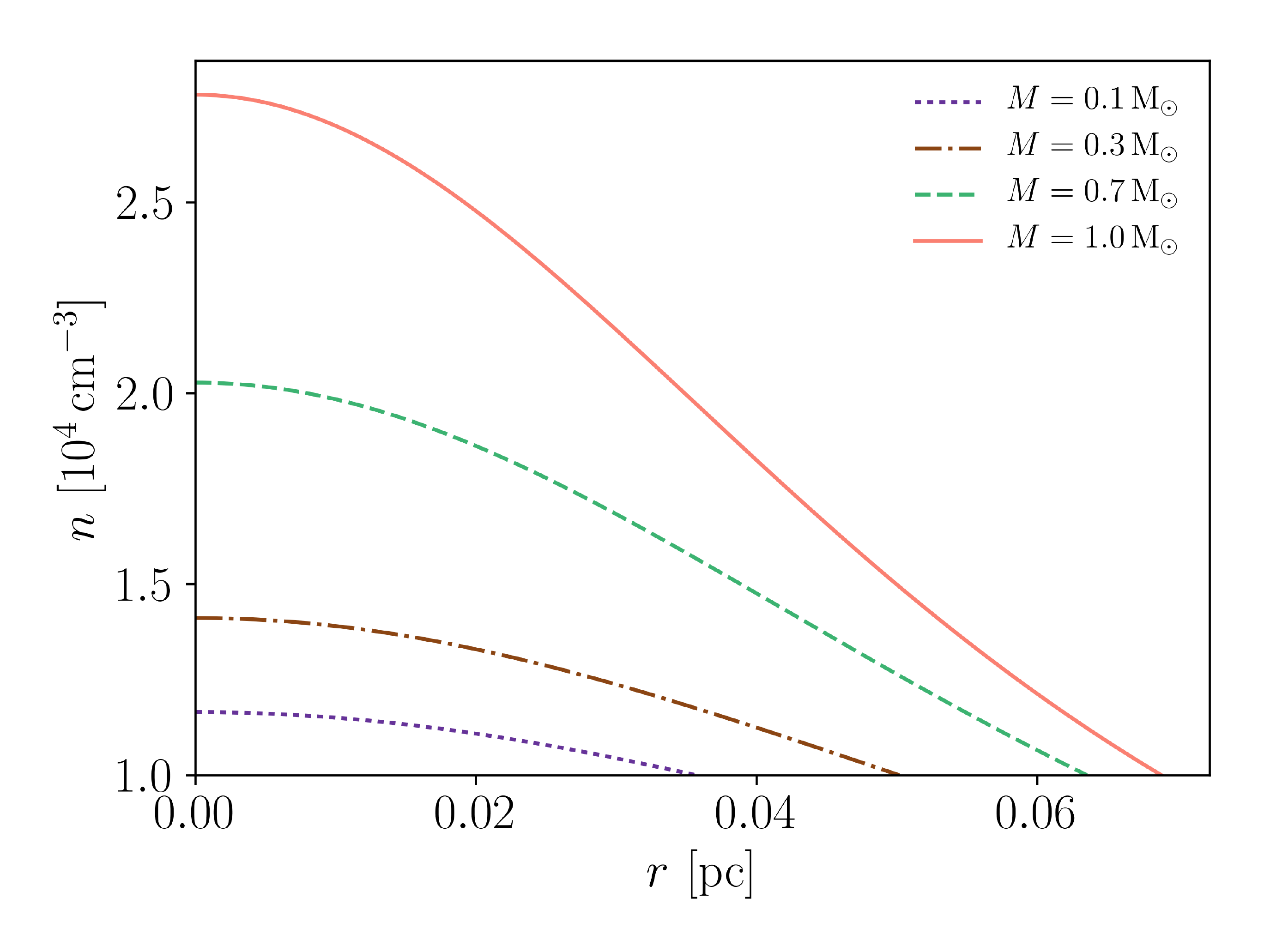}
\caption{Density profile of molecular spherical clumps at temperature $T_\h2=10$ K, confined by a medium with $T_\icm = 100$ K and $n_\icm = 10^3\, \cc$, computed with the BE model The outer density depends only on $T_\h2$ and the confining ICM pressure, while the mass sets the density at the centre of the clump.} 
\label{Fig1a}
\end{figure}

\subsection{Gaseous environment}
\label{icm}

\begin{table}
\centering
\begin{tabular}{ccccc}
\hline \hline
                                  & $n_\icm\,{\rm [\cc]}$	 & $T_\icm$ [K] & $T_\h2$ [K]              \\ \hline
\multicolumn{1}{c|}{Stellar case} & $10$        		 & $100 - 1000$  & $10$\\
\multicolumn{1}{c|}{Quasar case}  & $60$            		 & $2.2\times 10^7$ & $100$\\  \hline 
\end{tabular}
\caption{Clump and ICM properties at the onset of the photoevaporation process.}
\label{tabIC}
\end{table}

The gaseous environment where clumps are located plays a crucial role in determining properties such as temperature, density and confining pressure. We now describe the interclump medium (ICM) which surrounds clumps in the stellar and quasar case. The ICM properties are summarized in Tab. \ref{tabIC}.

We assume that clumps in a stellar surrounding are in pressure equilibrium with atomic gas, whose temperature depends on the distance and luminosity from the stellar source. We take $n_\textsc{at}\simeq 10^3 \,\cc$ as a typical density for this surrounding gas, and we compute the corresponding temperature according to the FUV flux, using the same method used to compute the temperature of the atomic phase in the clump (see Sec. \ref{shellstructure}). Such temperature ranges between $10^{2}$ K and $10^{3}$ K. The clump is exposed to the radiation of the massive star when it is engulfed in the growing HII region, for which the density is taken to be $n_\icm\simeq 10\,\cc$.

In the case of quasars, molecular clumps likely form from thermal instabilities within the outflow. Clumps detach from the hot phase at the discontinuity between the fast wind and the ISM \citep{Zubovas2014a}, starting from the distance at which the outflow has become energy-driven. This critical radius has been estimated by {\cite{Zubovas2012a}:
\begin{equation}
\label{criticalradius}
	R_c = 520 \, \sigma_{200} M_8^{1/2} v_{0.1} \, \mathrm{pc}\,\,,
\end{equation}
where $\sigma_{200}$ is the velocity dispersion in the host galaxy in units of 200 km ${\rm s}^{-1}$, $M_8$ is the mass of the SMBH in units of  $10^8\,\msun$ and $v_{0.1}$ is the wind velocity in units of 0.1 times the speed of light. In the outflow, the gas is heated to $T_\icm \simeq 2.2\times 10^7$ K and has a typical density $n_\icm \simeq 60$ cm$^{-3}$ \citep{Zubovas2014a, Costa2014}. The outflow fragments \citep{King2010, Nayakshin2012} because of thermal instabilties, so that one component cools to a low temperature. The existence of an equilibrium between a $10^4$ K and a $10^7$ K phase has been studied by \cite{Krolik1981}, while \cite{Zubovas2014a} show that an atomic clump requires a very short time to cool and turn to molecular form. Molecule formation can occur in the overdensities generated via thermal instabilities, since radiation is efficiently self-shielded and the gas deep into the clump is allowed to cool\footnote{This conclusion has been re-examined by \citet{Ferrara2016c}, who pointed out that molecule formation is problematic due to the efficient dust destruction by the outflow shock.}. When a clump starts to cool, we assume that it maintains pressure balance with the ICM until its temperature is $T\simgt 10^4$ K. Below such temperature the cooling time-scale is very short, and the evaporation process detailed in the next Sections happens before the clump can readjust to the external pressure. The final temperature of the molecular gas is about 100 K, in agreement with detections with CO and water vapour line emission \citep{Cicone2012a, Aalto2012b, Gonzalez2010}.

\subsection{Radiation field} 

Radiation affects the structure of a clump according to the shape of the emitted spectrum. In particular we are interested in ionizing (energy $h\nu \geq 13.6$ eV) and FUV photons ($6$ eV $< h \nu< 13.6 $ eV), whose flux $G_0$ is measured in units of the Habing flux\footnote{The Habing flux ($1.6 \times 10^{-3}\, {\rm erg}\, {\rm s}^{-1} {\rm cm}^{-2}$) is the average interstellar radiation field of our Galaxy in the range [6 eV, 13.6 eV] \citep{Habing1968}.}.

For stellar sources, we use black body spectra with different effective temperatures $T_{eff}$. In terms of solar luminosity ($\lsun$), OB stars have typical luminosities ranging between $10^3\lsun$ and $10^5\lsun$. Then, the effective temperature is given by
\begin{equation}
	T_{eff} = \left( \dfrac{L}{4\pi R_{\star}^2 \sigma_{\textsc{sb}}} \right)^{1/4}\,\,,
\end{equation}
where $\sigma_{\textsc{sb}}$ is the Stefan-Boltzmann constant, $R_{\star}$ is the star radius and $L$ is the bolometric luminosity. We compute $R_{\star}$ through the mass-luminosity and radius-luminosity relations by \cite{Demircan1991}, which for an OB star give
\begin{equation}
	R_{\star} = 1.33 \,\rsun \left( \dfrac{L}{1.02 \,\lsun} \right)^{0.142}\,\,,
\end{equation}
where $\rsun$ is the solar radius. Integrating the black body spectrum in the FUV band, typical values of the FUV flux are $G_0=10^2 - 10^4$ for gas at $0.3$ pc from sources with luminosities in the range $L=10^3 - 10^5\, \lsun$. In the same way, we integrate the spectrum for $h\nu \geq 13.6\,$ eV to obtain the ionizing flux.
	
In the case of quasars, the fundamental difference is the wide extension of the spectrum to the X-rays, so that ionizing radiation is much more intense in this case. An analytical expression for the ionizing flux can be found with the same approach as in \cite{Ferrara2016c}, thus obtaining that the specific (ionizing) luminosity for $\nu > \nu_L\simeq 3.3\times 10^{15}$ Hz is:
\begin{equation}
\label{quasarSED}
	L_{\nu} = 6.2\times 10^{-17}\left(\dfrac{\nu}{\nu_L}\right)^{\alpha-2}\,\left(\dfrac{L}{\mathrm{erg}\,{\rm s}^{-1}}\right)	\,{\rm erg}\,{\rm s}^{-1}\,{\rm Hz}^{-1}\,\,,
\end{equation}
where $\alpha= 0.5$ for a radio-quiet quasar \citep{Mortlock2011}. We assume eq. \ref{quasarSED} to be valid for energies below the cut-off value $E_c=300$ keV \citep{Sazonov2004, Yue2013}. Furthermore, we can easily infer a relation between the bolometric and FUV luminosity, setting $\nu = \nu_L$ in eq. \ref{quasarSED}. The spectrum is almost flat in the FUV band, with typical values $G_0\simeq 10^{3 - 5}$ at 1 kpc, for $L=10^{45-47}\,{\rm erg}\,{\rm s}^{-1}$.

We investigate the evolution of clumps irradiated by stars or quasars, with $L$ as the only free parameter determining the flux in the bands we are interested in.

\subsection{Clump structure} 

\begin{figure}
\centering
\includegraphics[width=0.495\textwidth]{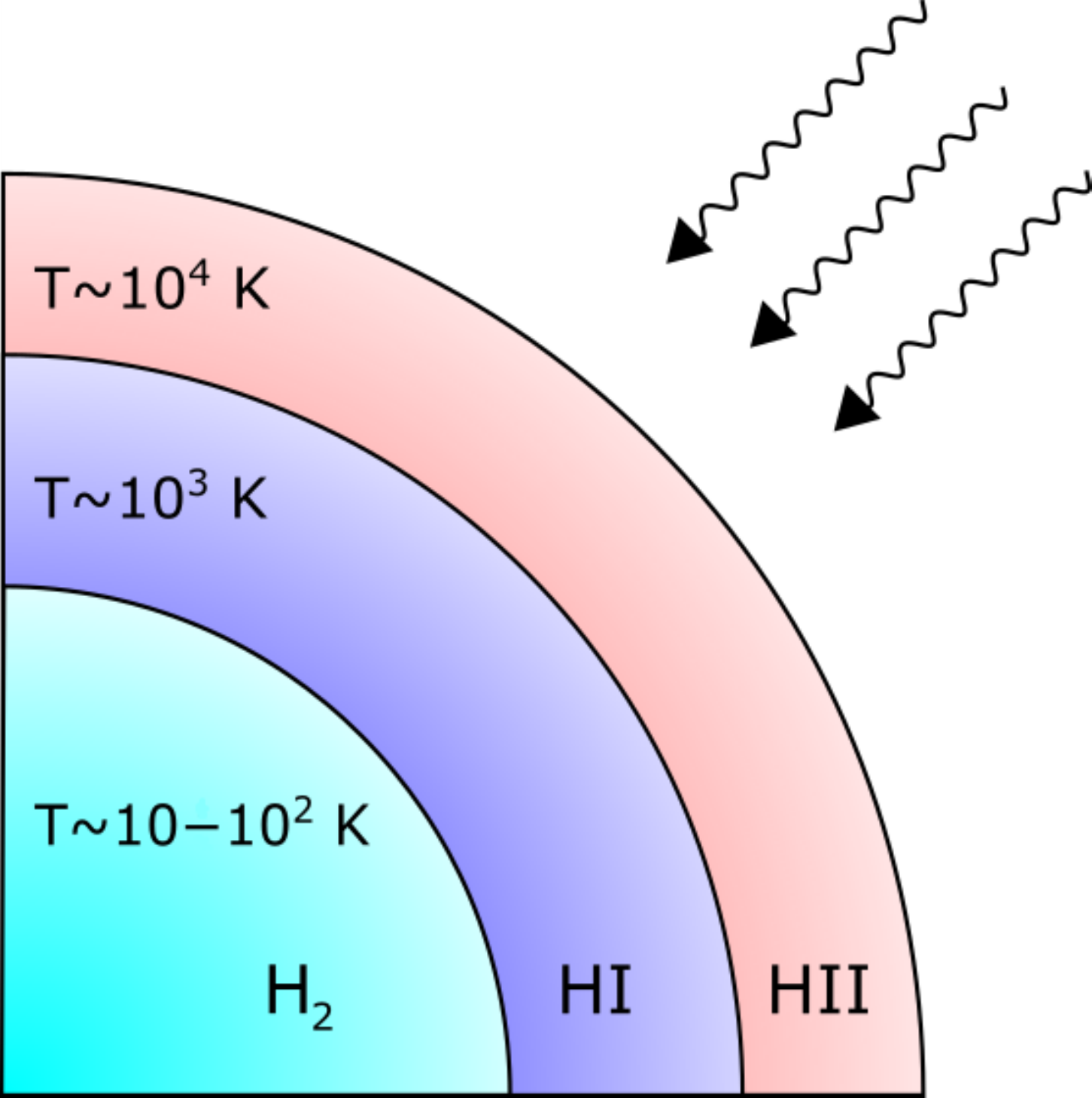}
\caption{Schematic structure of a clump exposed to UV radiation. In the sudden heating approximation, a molecular clump instantaneously develops a shell structure, with an ionized (HII) shell and a neutral (HI) shell, surrounding a cold and dense molecular core (H$_2$).} 
\label{Fig1}
\end{figure}

\begin{figure}
\centering
\includegraphics[width=0.47\textwidth]{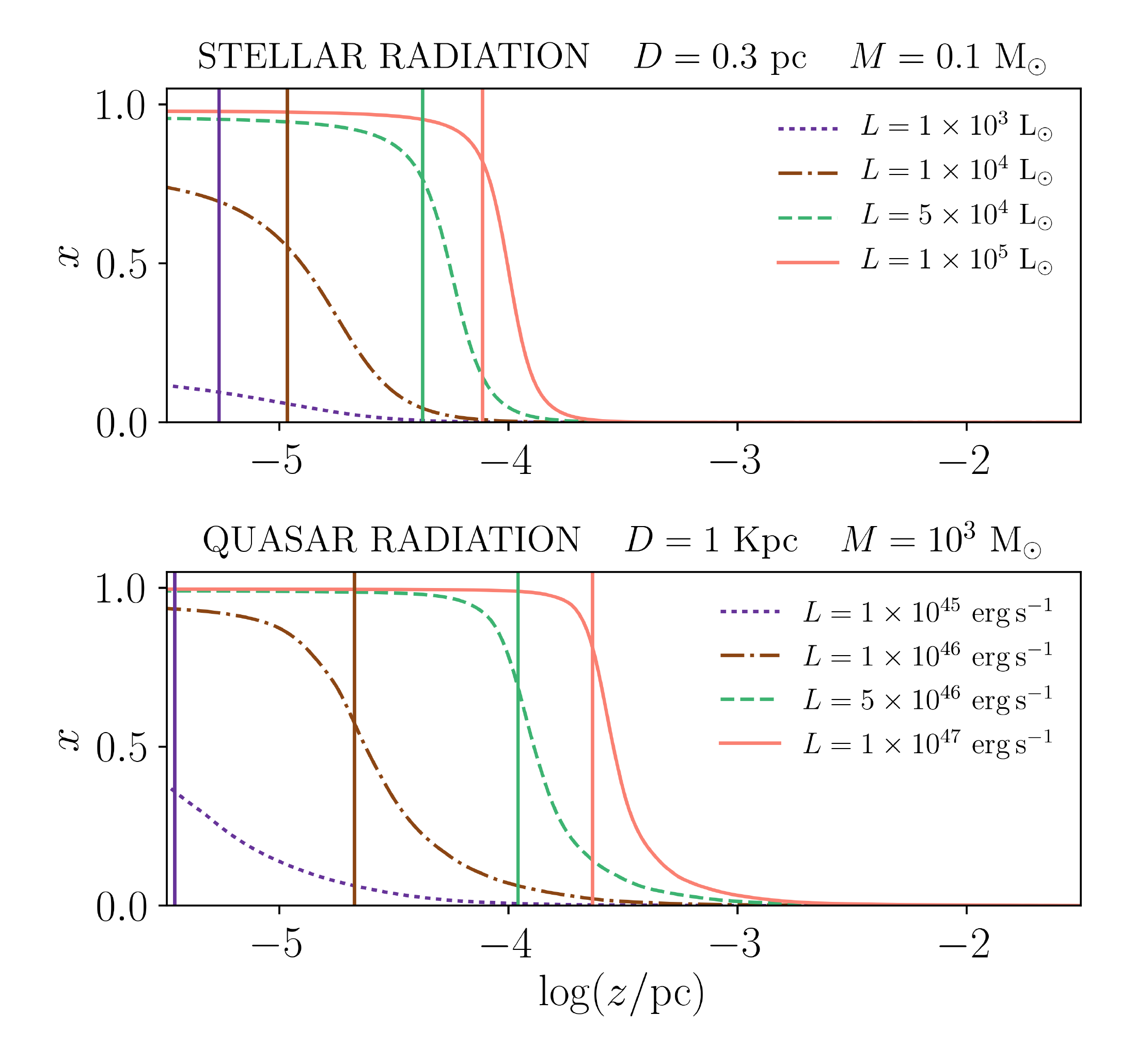}
\caption{The ionized fraction $x$ in the HII shell is plotted as a function of depth ($z=0$ is the surface of the clump) for different luminosities of the sources. Both the stellar case ({\bf upper} panel) and quasar case ({\bf lower} panel) are represented. Clumps have mass $0.1 \, \msun$ and $10^3 \, \msun$ for stars and quasars respectively, and their density in the HII shell is computed with a BE density profile. The confining ICM is as described in Sec. \ref{icm}. Vertical  lines mark the thickness of the HII shell computed as if the transition between ionized and neutral phase was a step function.} 
\label{Fig1b}
\end{figure}	

\label{shellstructure}
\begin{figure*}
\centering
\includegraphics[width=0.495\textwidth]{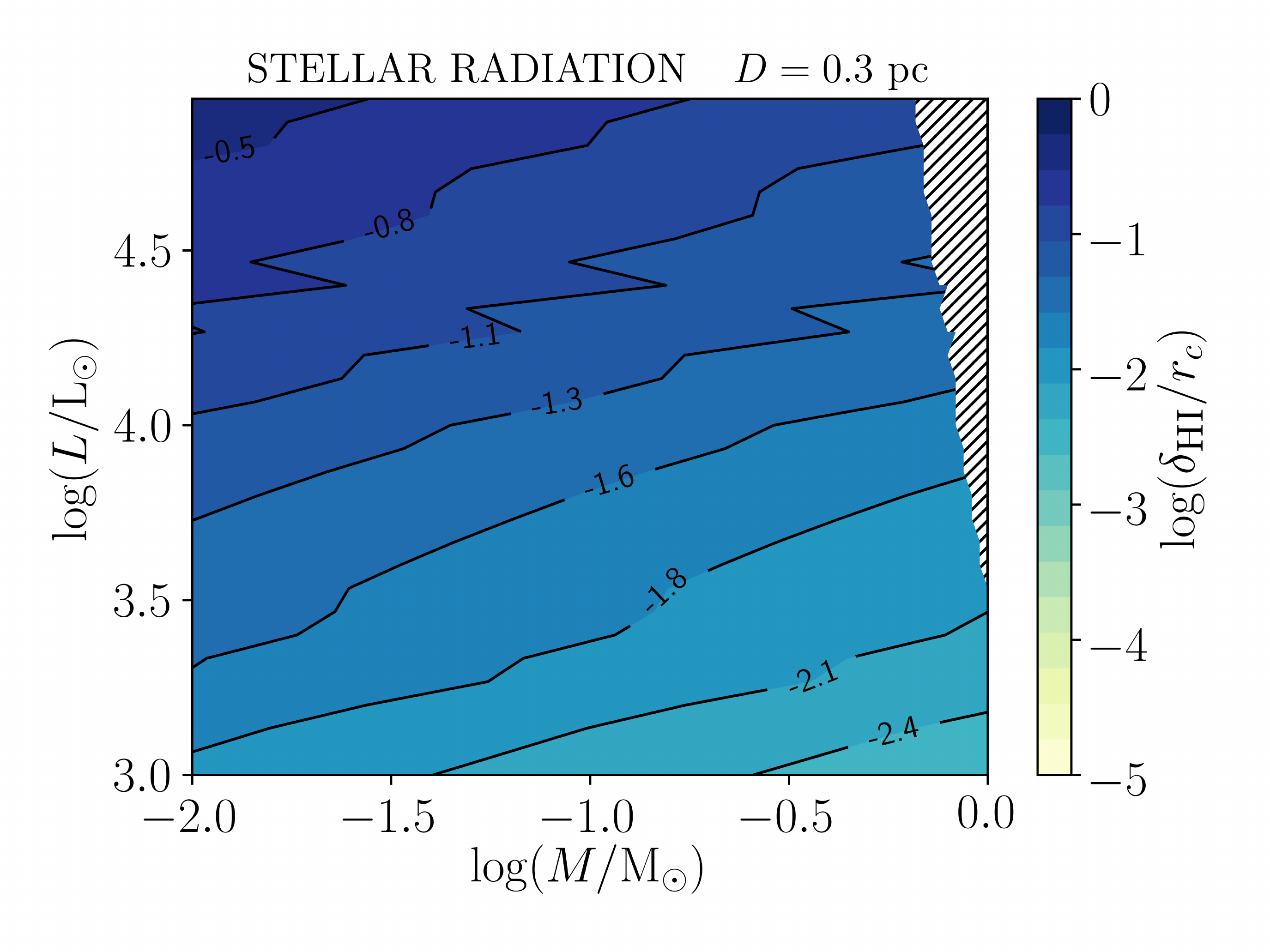}
\includegraphics[width=0.495\textwidth]{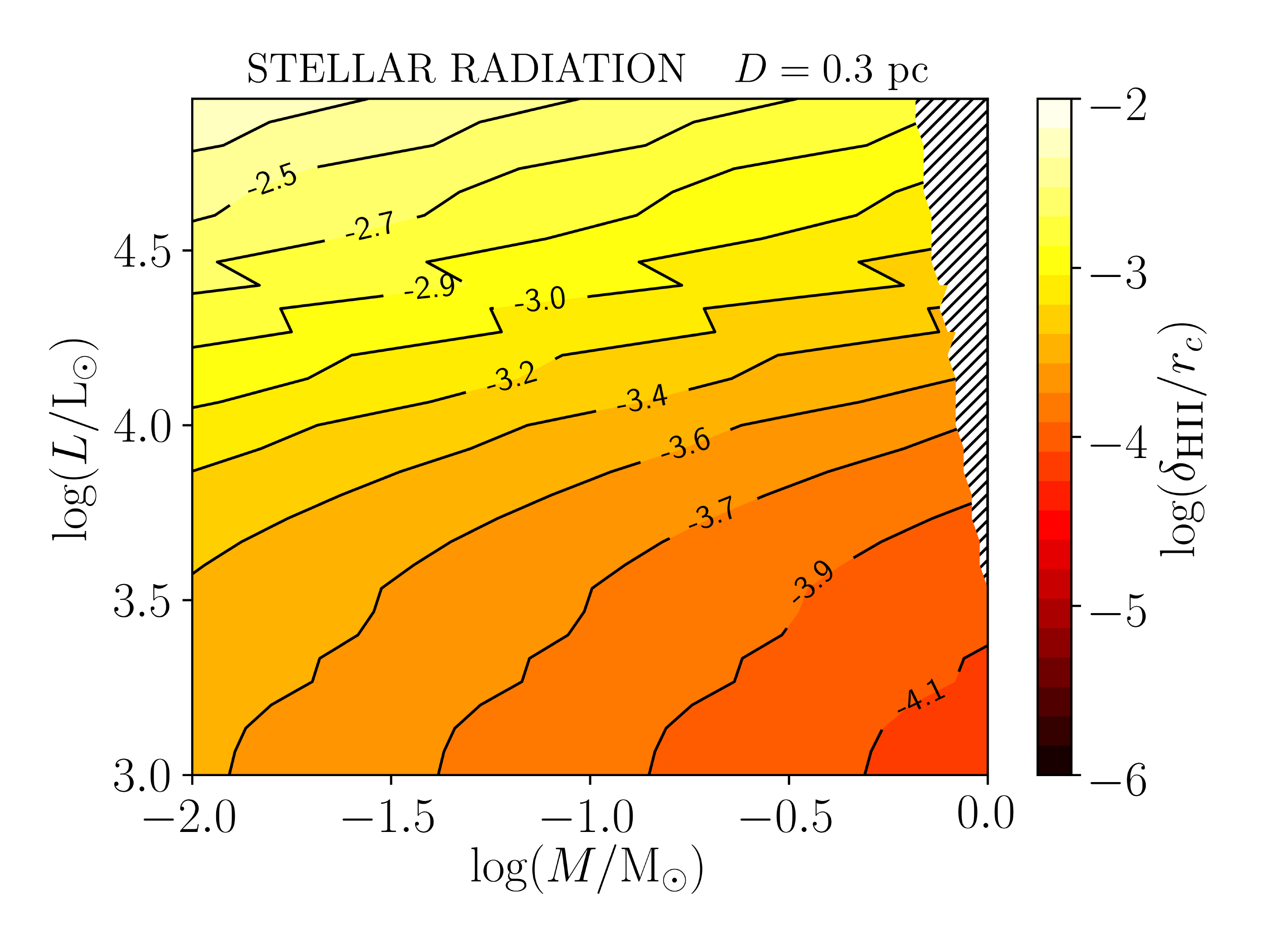}
\includegraphics[width=0.495\textwidth]{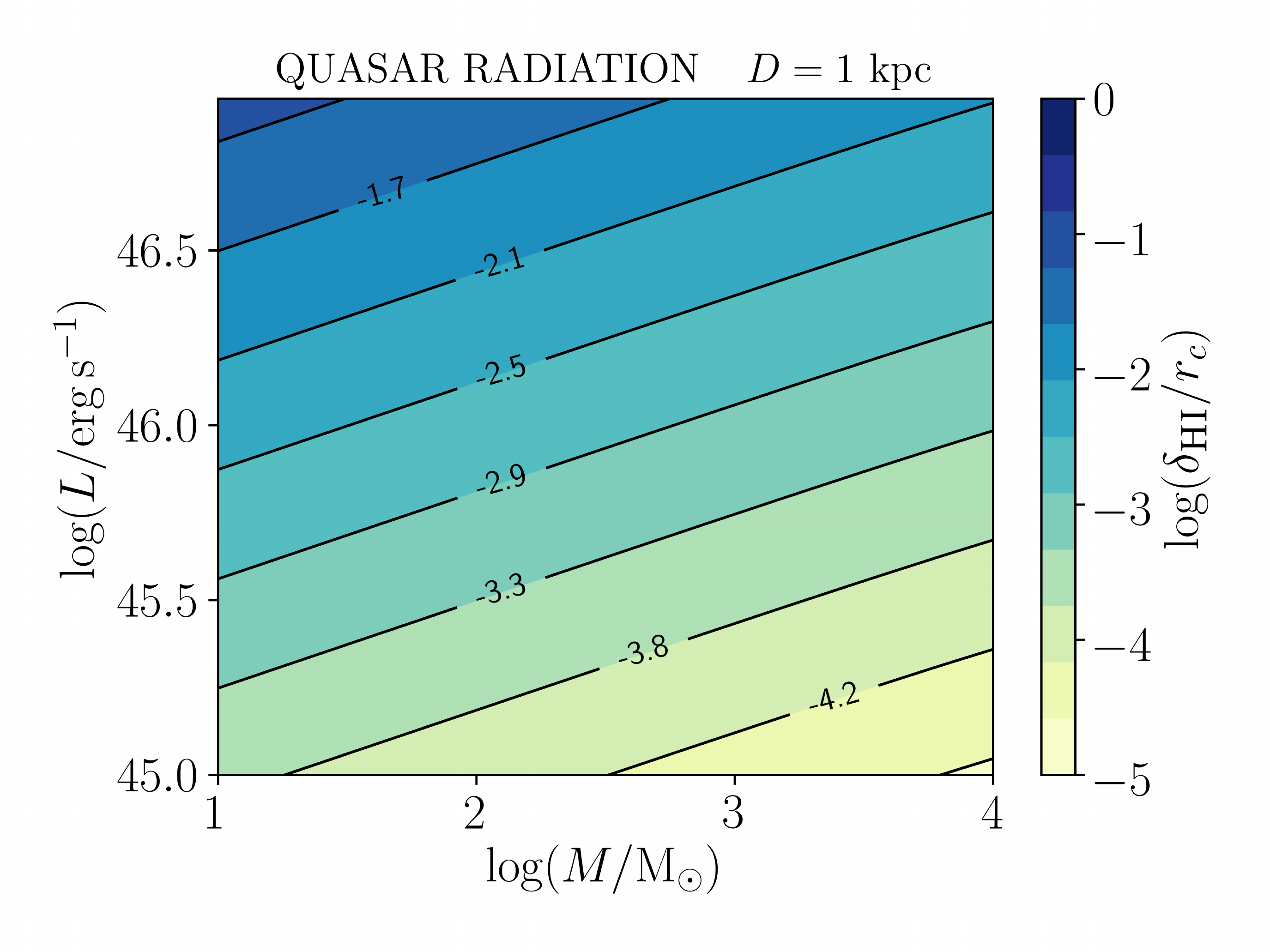}
\includegraphics[width=0.495\textwidth]{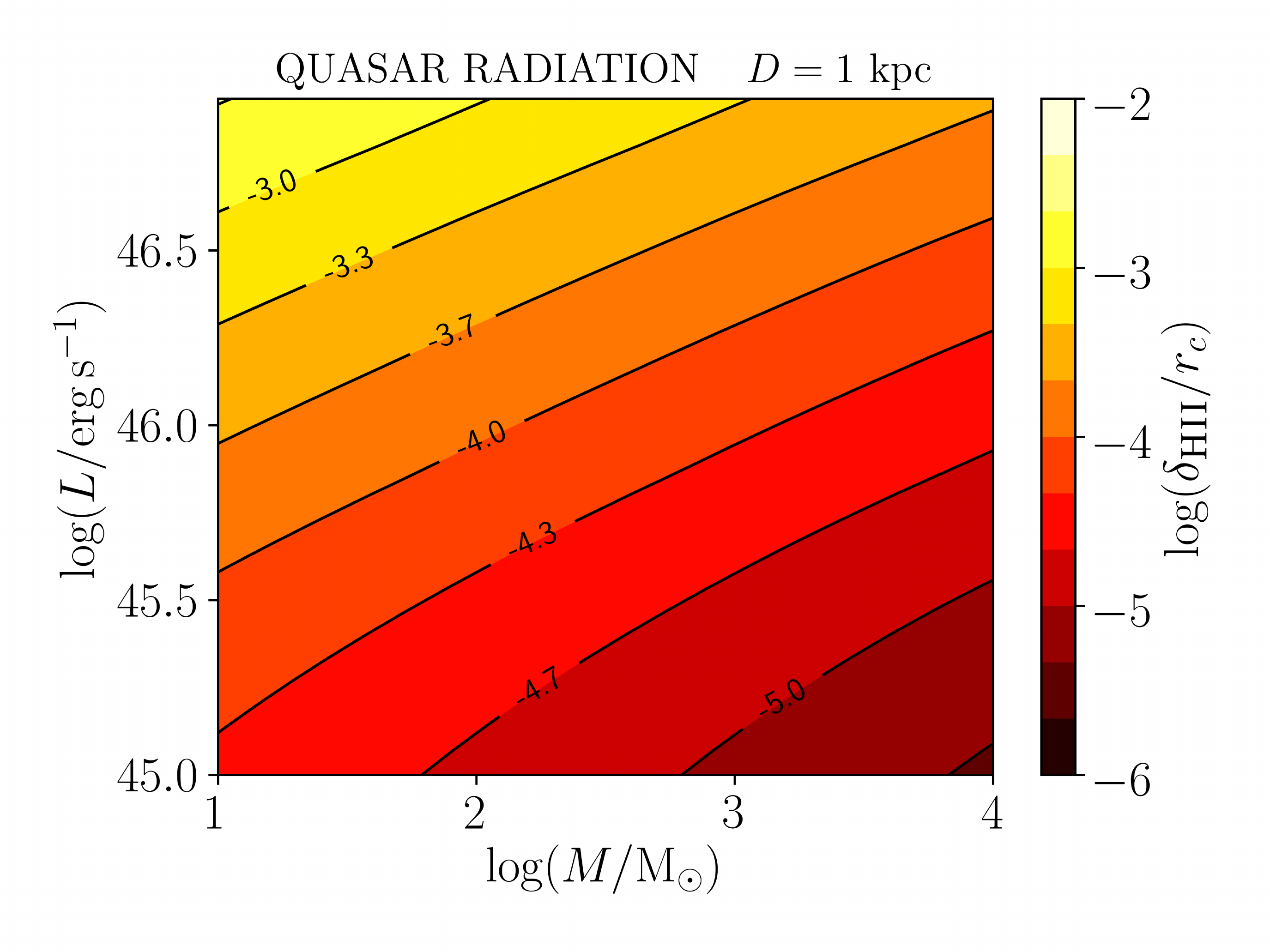}
\caption{Plots of the ratio of the shell thickness to the initial clump radius $r_c$, as a function of clump mass and source luminosity. {\bf Upper} panels: HI shell thickness $\delta_\hi$ ({\bf left}) and HII shell thickness $\delta_\hii$ ({\bf right}) when the clump is located in the surroundings of a massive star, at a distance of 0.3 pc. Clumps in the shaded region are not considered, since their mass is larger than the BE mass for collapse (eq. \ref{BEmass}). {\bf Lower} panels: HI shell thickness $\delta_\hi$ ({\bf left}) and HII shell thickness $\delta_\hii$ ({\bf right}) when the clump is illuminated by a quasar at a distance of 1 kpc. See the text for more details.}
\label{Fig11}
\end{figure*}

In this paper a clump is modeled as a Bonnor-Ebert (BE) sphere \citep{Ebert1955, Bonnor1956}, which is isotropically affected by an external impinging radiation field. Given the clump mass, the clump temperature and the confining pressure, the BE sphere model allows to compute the radial density profile inside the clump and its radius. In Fig. \ref{Fig1a} we show the density profile for clumps of different mass, with the same temperature $T=10$ K and a confining pressure $P=10^{-11}$ erg cm$^{-3}$ (corresponding to a confining medium with $T_\icm = 100$ K and $n_\icm = 10^3\, \cc$). The clumps have different radii and same outer density, set by the pressure equilibrium between the clump and the surrounding gas. The density increases towards the centre, with a steeper profile for larger values of the mass. A clump undergoes a collapse if its mass is larger than the BE mass:
\begin{equation}
\label{BEmass}
	M_\textsc{be} \simeq 1.18 \dfrac{c_s^4}{\sqrt{P_0 G^3}}\,\,,
\end{equation}
where $P_0$ is the confining pressure, $c_s$ is the isothermal sound speed and $G$ is the gravitational constant. We are considering only the thermal contribution to pressure, not accounting for turbulent and magnetic pressure. We underline that, apart from the use of a BE density profile, gravity is not included in the hydrodynamical equations for the clump evolution presented in Sec. \ref{shockdynamics}.

Given the clump density profile, we assume that the impinging radiation induces a shell-like structure, before any dynamical response of the gas to the photo-heating occurs (sudden heating approximation, see Fig. \ref{Fig1}). The FUV radiation is responsible for the formation of an atomic layer (HI shell). The more energetic part of the spectrum partially or totally ionizes an outer shell (HII shell), depending on the intensity of the source. This sets up the initial condition for the subsequent hydrodynamical evolution of the clump.

The sudden heating approximation means that radiation dissociates and ionizes particles and heats the gas to its final temperature instantaneously, while the clump shape is unaltered. This situation is often referred to as a R-type ionization front \citep{Spitzer1998}. To justify this assumption, we compare the sound-crossing time-scale $t_{cross}$ with the ionization timescale $t_i$ and the heating time-scale $t_h$. The Str\"{o}mgren theory adapted for plane geometry (which can be assumed when the radius of the clump is much smaller than the distance from the source) allows to compute the HII shell thickness, 
\begin{equation}
	\delta_\hii (t)=\delta_\hii (1-e^{{-n}\alpha_\textsc{b}t})\,\,,
\end{equation}
where $n$ is the gas number density, and $\alpha_\textsc{b}$ is the case B recombination coefficient  \citep[values in][]{Verner1996}. Thus, we have $t_i = 1/n\alpha_\textsc{b} $. On the other hand, the sound-crossing time-scale is $t_{cross} = r_c/c_s$ where $r_c$ is the clump radius and $c_s\sim \sqrt{k_\textsc{b}T/m_p}$ is the sound speed. Plugging in typical values, it is easily seen that the condition $t_i\ll t_{cross}$ is always satisfied for physically reasonable values of $r_c$ ($0.01- 1$ pc) and $c_s$ ($0.1 1$ km ${\rm s}^{-1}$ in the cold phase). Regarding the gas heating time-scale ($t_h$), a simple estimate gives us:
\begin{equation}
	\label{eq:heating time-scale}
 	t_h = \dfrac{k_\textsc{B} T_f }{\Gamma(T_f)}\,\,,
\end{equation}
where $k_\textsc{B}$ is the Boltzmann constant, $T_f$ is the final gas temperature and $\Gamma(T_f)$ is the heating rate (in erg/s), mainly due to photoionization. To give some examples, $\Gamma / n$ varies from $\sim 10^{-23}$ erg cm$^{3}$ s$^{-1}$ at a distance of 1 pc from a star with $L=10^3\,\lsun$, to $\sim 10^{-16}$ erg cm$^{3}$ s$^{-1}$ at 0.5 kpc from a quasar with $L=10^{47}$ erg s$^{-1}$ \citep[using the heating function by][]{Gnedin2012}. The result is that for any value of $r_c$ and $c_s$ of interest, the condition $t_h\ll t_{cross}$ holds.
	
In what follows, we discuss how we compute the thickness and the temperature of each shell in the clump. The thickness of the HI shell ($\delta_\hi$) is defined as the depth at which hydrogen is found in molecular form. \cite{Tielens1985} find how such depth (expressed as hydrogen column density) scales with the gas density and the FUV flux:
\begin{equation}
\label{columndensity}
	N_\textsc{H} \propto n^{-4/3}G_0^{4/3}\,\,,
\end{equation}
and that the thickness for $n=2.3 \times 10^5$ cm$^{-3}$ and $G_0=10^5$ is $\delta_\hi =1.5\times 10^{16}$ cm. Then from eq. \ref{columndensity} we find for the shell thickness
\begin{equation}
	\label{deltaHI}
	\delta_\hi  = 0.034 \, \left(\dfrac{n}{10^5\,{\rm cm}^{-3}}\right)^{-7/3}\, {\left( \dfrac{G_0}{10^5}\right)}^{4/3}\, {\rm pc}\,\,.
\end{equation}
\cite{Kaufman1999} outline that the PDR temperature is rather constant before it drops to the low values of the molecular core. They plot the temperature for different values of the density and the FUV flux, and we use a fit of their model to estimate the temperature of the HI shell. 

The outer shell presents a partial or total ionization, depending on its density and the intensity of the impinging radiation field. We compute the equilibrium temperature as a function of depth into the shell by balancing photoionization heating and  recombination cooling, line cooling and bremsstrahlung. The presence of a radiation field alters the heating and cooling rate: 1) the ionized fraction of each species is modified and thus the cooling rate by line emission is changed accordingly; 2) injection of photoionized electrons in the gas provides an extra heating term. Approximate heating and cooling functions, assuming collisional equilibrium but non-zero radiation field, are provided by \cite{Gnedin2012}. We assume a fiducial value for metallicity, i.e. the mass fraction of elements heavier than helium, of $Z=0.02$ \citep[close to the solar value from][]{Anders1989}, noticing that metals are important for the energetics of the gas, but their contribution to its dynamics (determined by gas pressure) is negligible. Moreover, we account for Compton heating, which we expect to be important for hard radiation fields:
\begin{equation}
	H_C = \dfrac{\sigma_T F}{m_e c^2} (\left\langle h\nu \right\rangle - 4k_\textsc{b}T)\,\,,
\end{equation}
where $\sigma_T$ is the Thomson cross section, $m_e$ is the electron mass, $F$ is the total flux and $\left\langle h\nu \right\rangle$ is the average photon energy beyond the Lyman limit.

Once the temperature profile is computed, we obtain the ionization profile by balancing photoionization, collisional ionization and recombination:
\begin{equation}
	\gamma(T) n_e n_p + n_\textsc{h} \int_{\nu_L}^\infty \dfrac{F_\nu}{h\nu}e^{-\tau_\nu}a_\nu(T) d\nu = \alpha_\textsc{b}(T) n_e n_p \,\,,
\end{equation}
where $n_e$, $n_p$ and $n_\textsc{h}$ are the electron, proton and neutral hydrogen density, respectively; $F_\nu$ is the specific flux from the source, $\tau_\nu$ is the optical depth, $a_\nu$ and $\gamma$ are the photoionization cross section and the collision ionization coefficient, respectively \citep[analytical fits by][]{Verner1995, Cen1992}. In Fig. \ref{Fig1b} we plot the ionization fraction $x=n_e/n=n_p/n$, as a function of the depth into the clump, for different source luminosities. The ionization profile varies smoothly throughout the HII shell, between the edge of the clump and the PDR, both in the stellar and quasar case. Nevertheless, the region where $x$ is varying is of the order of $10^{-5}$ pc, which is negligible with respect to typical clump radii ($0.01 - 1$ pc). Then we adopt a reference value for the HII shell thickness $\delta_\hii$, computed with the approximation of a sharp boundary between ionized and phase, in the same way as done for the Str\"{o}mgren radius for a stellar HII region. Further assuming that the clump radius is much smaller than the distance from the source and that it is illuminated isotropically, the HII shell depth is
\begin{equation}
	\delta_\hii = \dfrac{1}{x^2_{max}n^2 \alpha_\textsc{b}(T)} \int_{\nu_L}^\infty \dfrac{F_\nu}{h \nu}\, \mathrm{d}\nu\,\,,
\end{equation}
where $x_{max}$ is the maximum ionization fraction, at the edge of the clump. The thickness $\delta_\hii$ is shown with dashed lines in Fig. \ref{Fig1b}.

We restrict our analysis to clumps where the shell thickness is much smaller than the molecular core radius. This allows us to determine the densities of the HI and the HII shells by using the outer density of the BE sphere. Furthermore, it simplifies our calculations, because we can consider separately the evolution of the shells and the core, since the dynamical time-scale of the former  is much shorter than the core one. In Fig. \ref{Fig11} we plot the ratio of the HI shell thickness (left panels) and the HII shell thickness (right panels) to the total clump radius $r_c$ as a function of the clump mass, for stars and quasars with different luminosities. The atomic shell is always thicker than the ionized shell, showing a large self-shielding effect of the ionized gas. The typical masses of clumps are different for stars and quasars, and we consider only masses smaller than the BE mass for collapse. The distances of clumps from the source are fixed in the two scenarios, and are reasonable for molecular gas engulfed by an expanding stellar HII regions (the distance scale is given by the Str\"{o}mgren radius) and clumps forming in quasar outflows (critical radius given in eq. \ref{criticalradius}).

Clumps presenting a ratio $\delta_\hi /r_c=1$ are completely dissociated on a time-scale $t_i$, and the analysis of this paper restricts to clumps where $\delta_\hi \ll r_c$. From the plot, we see that such condition is usually verified and breaks only in the stellar case for small clumps ($M <
 0.01\,\msun$) and very intense sources ($L\simeq 10^5\, \lsun)$.

\subsection{Shock dynamics inside the clump} 
\label{shockdynamics}

\begin{figure}
\centering
\includegraphics[width=0.47\textwidth]{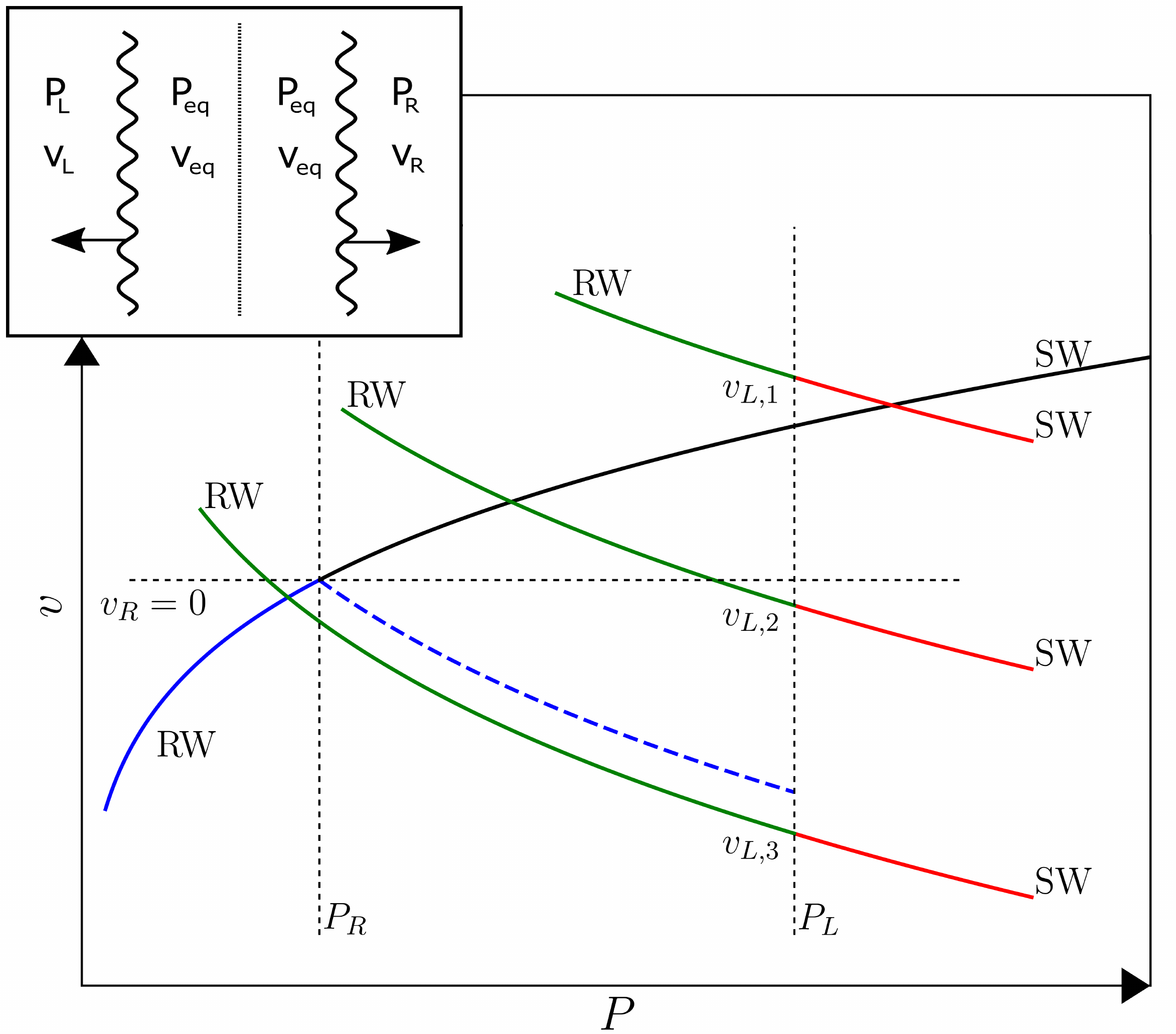}
\caption{
The sketch in the upper right corner shows the general arbitrary discontinuity problem: a gas has initial conditions ($P_\textsc{l}$, $v_\textsc{l}$) and ($P_\textsc{r}$, $v_\textsc{r}$) to the right and to the left of the discontinuity (dotted line) respectively. Wavy lines are shock waves (SW) or rarefaction waves (RW) originating from it, in order to set a continuous value $P_{eq}$ and $v_{eq}$ of pressure and velocity. The plot shows the qualitative solution of the problem with $P_\textsc{l}>P_\textsc{r}$ and $v_\textsc{r}=0$. The blue-black line connects all the possible final states on the right side, achieved through a SW or a RW. Similarly, the green-red lines the possible final states to the left side, for different values of $v_\textsc{l}$. The intersection of the lines for the regions to the left and to the right gives the solution, and different types of waves (shock or rarefaction waves) are required to get to the final state, according to the value of $v_\textsc{l}$.}
\label{Fig2}
\end{figure}

\begin{figure}
\centering
\includegraphics[width=0.495\textwidth]{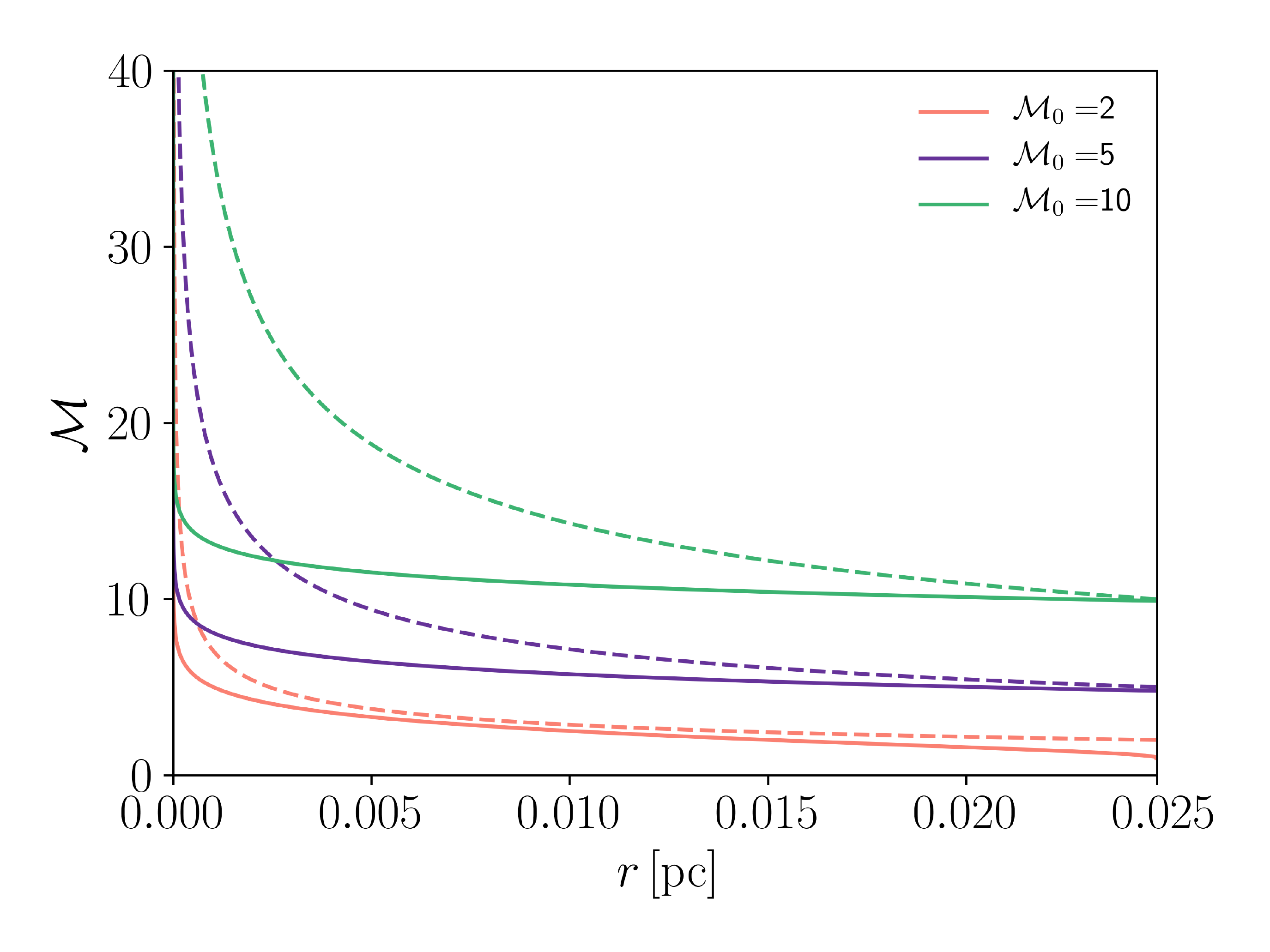}
\caption{Mach number radial profile, $\mathcal{M} (r)$, for different initial Mach numbers $\mathcal{M}_0$ of isothermal shocks at the clump surface, computed considering the density variation in the clump ({\bf solid} line) according to eq. \ref{sphericalshock} or using the power law of eq. \ref{guderley} ({\bf dashed} line) valid for homogeneous density and adiabatic shocks \citep[analytical solution by][]{Guderley1942}. The clump considered in the plot has mass $0.1\, \msun$ and radius $0.025$ pc, and it is made of molecular gas (so that the power law index is $n(\gamma)\simeq 0.394$).}  
\label{Fig4}
\end{figure}

Having set the initial conditions on a clump, i.e. a core-double shell structure, now we can study its dynamical evolution for $t>0$. The different layers in the clumps have different pressures, so that a shock or rarefaction waves originate, enforcing a continuous value of pressure and velocity across the contact discontinuity between two layers. 

The cooling time-scale of a gas at temperature $T$ is
\begin{equation}
	\label{eq:cooling time-scale}
 	t_{cool} = \dfrac{k_\textsc{B} T}{\Lambda(T)}\,\,,
\end{equation}
with $\Lambda$ being the cooling function given by \citet{Neufeld1995} for molecular gas  and by \citet{Tielens1985} for PDRs. For the range of temperatures and $n$, $r_c$ and $c_s$ values of interest here, $t_{cool} \ll t_{cross}$. Thus the fluid motion and the propagation of any disturbance in the gas (as shock and rarefaction waves) can be safely considered as isothermal processes. 

A qualitative diagram of the possible outcomes at an arbitrary discontinuity is shown in Fig. \ref{Fig2}. In the situation considered in the upper inset, a gas has a pressure $P_\textsc{r}$ to the right of an interface, and a pressure $P_\textsc{l}$ to the left (with $P_\textsc{l}>P_\textsc{r}$). The velocity to the right is $v_\textsc{r}=0$, while we consider different values $v_\textsc{l,1}$, $v_\textsc{l,2}$ and $v_\textsc{l,3}$ for the velocity to the left. The solid lines connect to the initial state all the possible final states of the gas, when it is crossed by a rarefaction wave (RW) or a shock wave (SW). For example, the points on the blue line represent the possible final states of the gas to the right when it is crossed by a rarefaction wave.  The solution of the discontinuity problem is obtained when the lines departing from the two initial states of the gas to the left and to the right intersect, since the final values of $P$ and $v$ must be the same across the discontinuity. This also determines which kind of wave is required, i.e. a SW or a RW. 

The solution of the problem for given values of the initial pressure, density and velocity across the discontinuity is obtained numerically, imposing the final pressure and velocity to be continuous. The post-shock values are obtained solving the isothermal Rankine-Hugoniot conditions \citep{Rankine1870}
\begin{subequations}\label{eq:flowequations}
\begin{align}
\rho_{0}v_{0}          &=\rho_{1}v_{1}\\
\rho_{0}v_{0}^{2}+P_{0}&=\rho_{1}v_{1}^{2}+P_{1}\\
T_0 &= T_1
\end{align}
\end{subequations}
where the subscript 0 is used for pre-shock values and the subscript 1 for post-shock values, with $v$ velocity in the shock front frame. Rearranging the equations \ref{eq:flowequations}, it is possible to write the following relations
\begin{subequations}\label{eq:postshockvalues}
\begin{align}
\rho_{1}          &=\rho_{0}\mathcal{M}^2\,\\
P_{1}          &=P_{0}\mathcal{M}^2\,\\
v_{1}          &=v_{0}/\mathcal{M}^2\,
\end{align}
\end{subequations}
with $\mathcal{M} = v_{0}/c_s$ being the shock Mach number. 

On the other hand, rarefaction waves are not discontinuities and values of flow variables across such waves are obtained following \cite{Zeldovich2002} and adapting the calculations to the isothermal case. Consider a wave originating at $x=0$ and propagating toward $x>0$, such that the final velocity after the wave has completely passed is $v_f=-U$. The profile between the \quotes{wave head}, moving at the initial sound speed $c_{s,0}$ in the gas, and the \quotes{wave tail}, moving at speed $v_{\mathrm{tail}}=c_{s,0} - ({\gamma+1})U/2$, is 
\begin{subequations}
\label{rarefactionprofile}
\begin{align}
v(x)		&=-\left(c_{s,0}-x/t \right)\\
\rho(x)	&=\rho_{0}\exp\left(x/c_{s,0}t-1 \right)\\
P(x)		&=P_{0}\exp\left(x/c_{s,0}t-1 \right)
\end{align}
\end{subequations}
where $\rho_{0}$ and $P_{0}$ are the values of density and pressure before the rarefaction has passed.

Since shock waves are discontinuities, an interaction between two shocks can be treated as an arbitrary discontinuity between post-shock values of flow variables. To simplify our analysis, we also consider interactions involving rarefactions as discontinuities, by accounting only for the post-rarefaction values of flow variables. In App. \ref{RWcollision} we compare this approach with a numerical solution of the fluid dynamics equations, showing that the two results differ negligibly.

To compute the shock speed inside the clump, we have to account for the spherical geometry and for the density gradient given by the BE profile. Following \cite{Whitham1958}, the flow equations can be written as
\begin{subequations}
\label{wavepropagation}
\begin{align}
	\partial_t \rho + \partial_r(\rho v) + \rho v \dfrac{\partial_r A(r)}{A(r)} &= 0 \label{wavepropagation1}\\
	\partial_t v + v \partial_r v + \dfrac{1}{\rho} \partial_r P - \dfrac{1}{\rho_0(r)} \partial_r P_0(r) &=0 \label{wavepropagation2}
\end{align}
\end{subequations}
where $r$ is the radial coordinate, $A(r)=4\pi r^2$ in the spherical case, $\rho_0(r)$ and $P_0(r)$ are the initial density and pressure profiles for a BE sphere. Eq. \ref{wavepropagation1} and eq. \ref{wavepropagation2} can be combined to give the equivalent equation valid along the curves $dr/dt = r+c_s$ in the ($r,t$) plane (called the $C_+$ characteristics):
\begin{equation}
	\mathrm{d}P + \rho c_s \mathrm{d} v + \dfrac{\rho c_s^2 v}{v+c}\dfrac{A'(r)}{A(r)} - \dfrac{\rho c_s}{v + c_s}\dfrac{1}{\rho_0(r)} P'_0(r)=0\,\,,
\label{eq:characteristiceq}
\end{equation}
where the prime denotes the derivative with respect to $r$. 
According to \cite{Whitham1958} the shock trajectory in the ($r,t$) plane is approximately a $C_+$ characteristic, so that eq. \ref{eq:characteristiceq} can be applied along the shock. Then we can write eq. \ref{eq:characteristiceq} as a function of the Mach number $\mathcal{M}$, substituting the post-shock values $P$, $\rho$, $v$ from eq. \ref{eq:postshockvalues}:
\begin{equation}
	\dfrac{d\mathcal{M}}{dr} = - \dfrac{1}{2}  \dfrac{\mathcal{M}^2}{\mathcal{M}^2 - 1} \frac{A'(r)}{A(r)} + \dfrac{1}{2} \dfrac{\mathcal{M}^3}{\mathcal{M}+1} \dfrac{P_0'(r)}{P_0(r)}\,\,,
	\label{sphericalshock}
\end{equation}
which is a differential equation for $\mathcal{M}$ as a function of $r$. 

In Fig. \ref{Fig4}, the solid lines show the the numerical solution of eq. \ref{sphericalshock} for a molecular clump with mass $M=0.1\,\msun$ and radius $r_c=0.025\,{\rm pc}$, for different values of the initial Mach numbers $\mathcal{M}_0$ of the shock at the edge of the clump, assuming an isothermal shock. For comparison, the dashed line is the classical analytical solution obtained by \citet{Guderley1942} for a homogeneous density distribution, and in the limit of a strong adiabatic shock:
\begin{equation}
	\mathcal{M}(r) = \mathcal{M}_0 \left(\dfrac{r_0}{r} \right)^{n(\gamma)}\,\,,
	\label{guderley}
\end{equation}
where $r_0$ is the radius of the bubble, and $n(\gamma)$ is an exponent depending on the adiabatic coefficient $\gamma$ (e.g. $n(5/3)\simeq 0.543$ for monoatomic gas and $n(7/5)\simeq 0.394$ for diatomic gas). As opposed to Guderley solution, the isothermal shock speeds up considerably only at a smaller radius. After the shock wave has reached the centre, a reflected shock will travel outwards. The velocity as a function of radius has the same profile of the focusing shock.

\begin{figure*}
\centering
\includegraphics[width=0.99\textwidth]{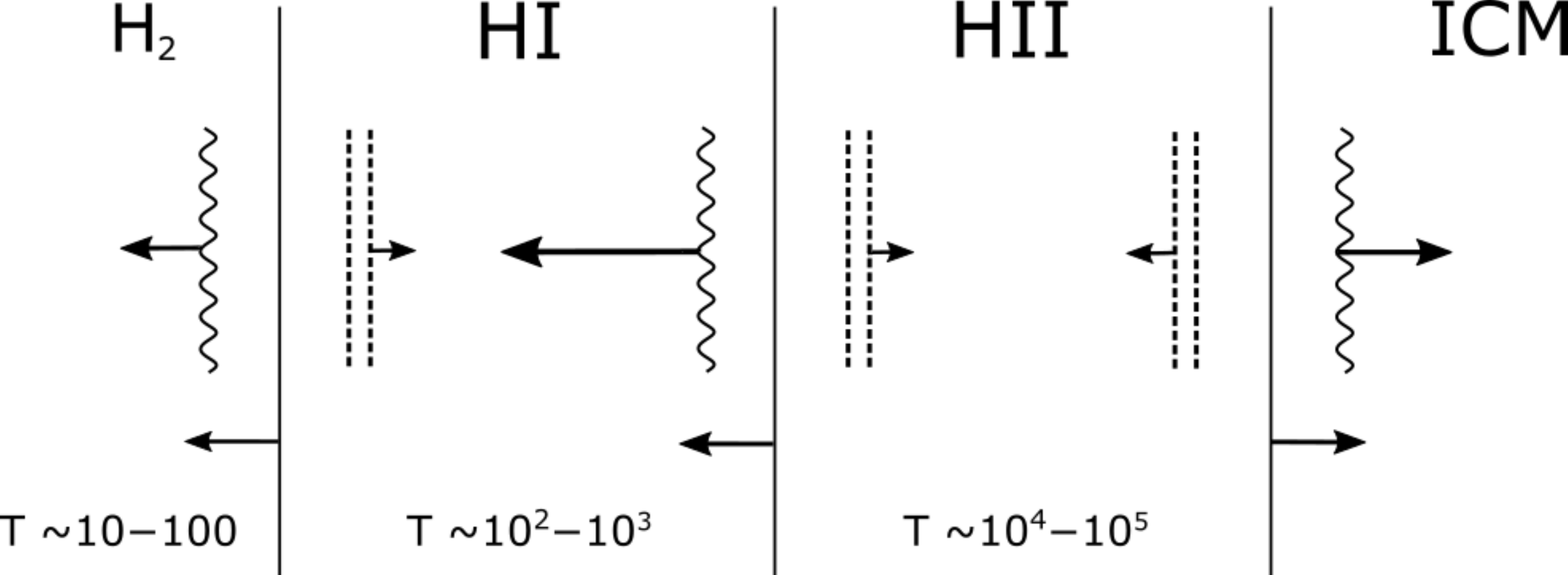}
\caption{Schematic representation of waves propagating in a clump suddenly heated by radiation, where wavy lines are shock waves and double dashed lines are rarefaction waves. High pressure shells drive a shock into adjacent inner shells, and as a result a rarefaction wave propagates back. Discontinuity interfaces move in the same direction of shock waves, at the post-shock speed. The result is an expansion of the two shells and a contraction of the core. See text for the a detailed description.} 
\label{Fig3}
\end{figure*}


\section{Results}
\label{Res}

\begin{figure*}
\centering
\includegraphics[width=0.495\textwidth]{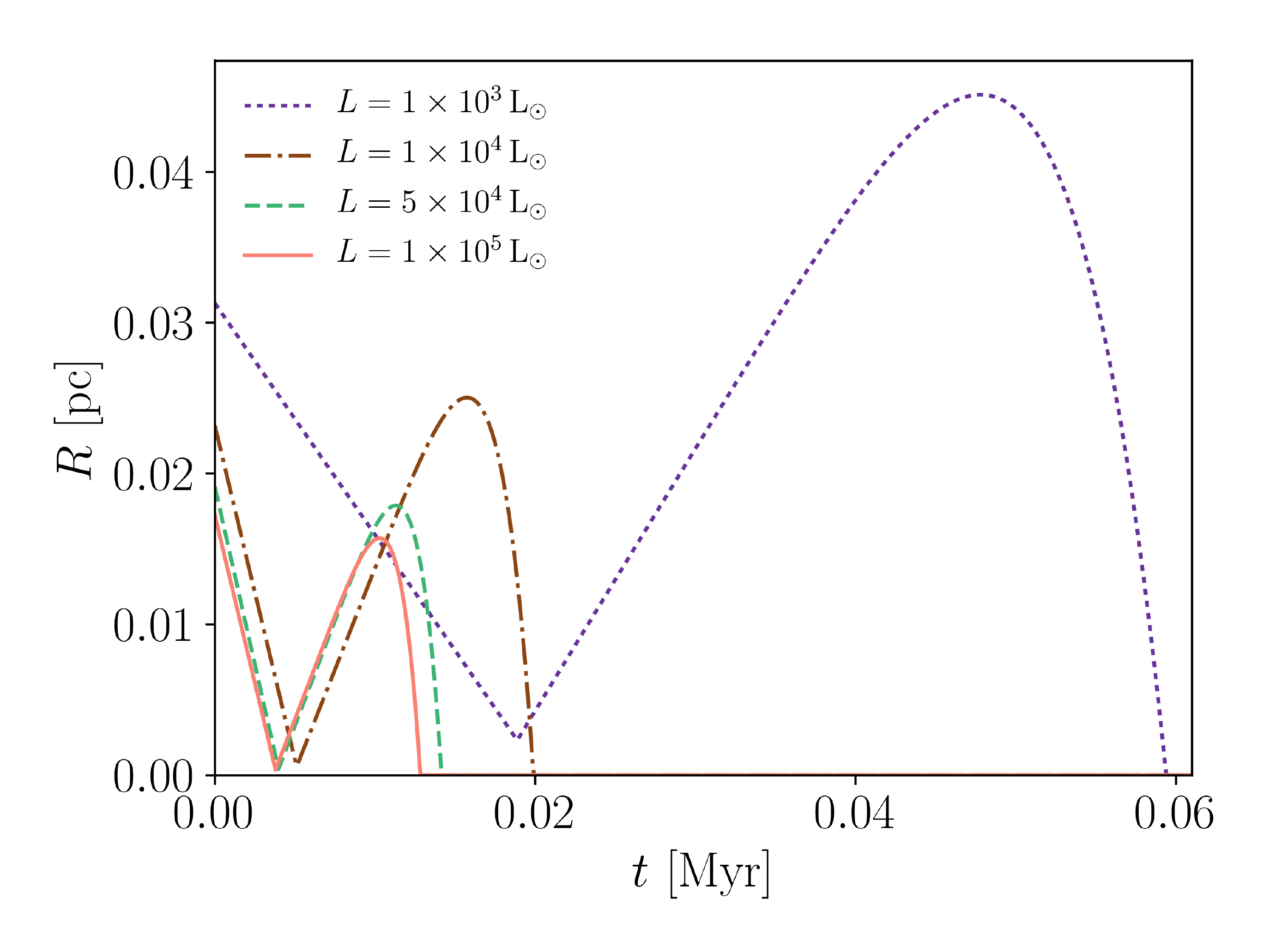}
\includegraphics[width=0.495\textwidth]{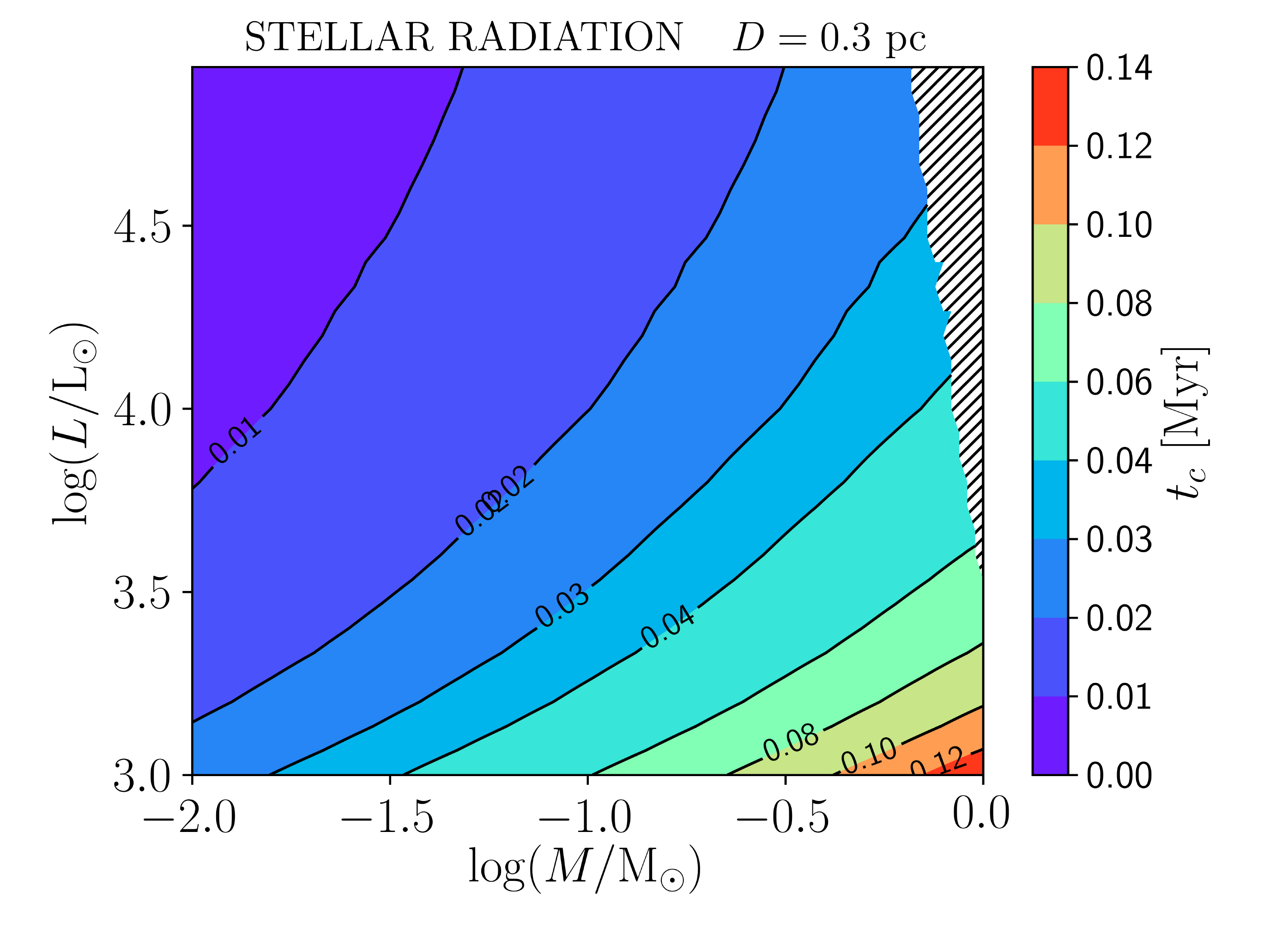}
\caption{{\bf Left}: radius, $R$, of the molecular core of a clump as a function of time, when the clump is exposed to stellar radiation of different luminosity $L$. The clump has a mass of $0.1\, \msun$, initial radius $r_c \simeq 0.02\,{\rm pc}$,  and is located 0.3 pc away from the source. {\bf Right}: Lifetime ($t_c$) of the clumps (located at 0.3 pc from the source) as a function of their mass and stellar source luminosity. Clumps in the shaded region are not considered, since their mass is larger than the BE mass for collapse (eq. \ref{BEmass}). }
\label{Fig5}
\end{figure*}

In our model, a clump exposed to UV radiation develops an ionized PDR (iPDR) at its surface. First, we inspect the qualitative behaviour of the structure simply using the arbitrary discontinuity criterion. Fig. \ref{Fig3} shows a diagram of shock and rarefaction waves propagating inside the clump because of the pressure difference between adjacent layers.

The HII shell pressure ($P_\hii $) is higher both than the pressure of the HI shell ($P_\hi $) and the pressure of the confining ICM ($P_\icm $), hence two rarefaction waves cross the HII shell, originating from its edges. Since $\delta_\hii\ll r_c$, the evolution of the HII shell has a much shorter time-scale than the clump evaporation time. The rarefaction waves which propagate into it interact and reflect at its edges, determining a complex density profile. Nevertheless, the global effect is that the HII shell expands decreasing its density, eventually becoming completely transparent to the ionizing radiation (i.e. the mean free path of photons is much larger than the shell thickness) }. 

On the other hand, $P_\hi $ is lower than $P_\hii $, but higher than $P_{\textsc{h}_2}$. Thus, a shock is driven from the HII shell into the HI shell, and a rarefaction wave propagates from the discontinuity with the core (see Fig. \ref{Fig3}). Once the shock has crossed the HI shell, it reaches the core surface and speeds up its contraction. As a result, the inner boundary of the HI shell moves faster than the outer boundary, and the HI shell is also expanding and becoming transparent on a time-scale shorter than the core evolution time-scale.
 
The cold ($T_\h2 \simeq 10- 100$ K) molecular core is compressed because of the shock wave originating at the discontinuity with the atomic shell and propagating towards the centre. In addition, the shock wave originating at the HII/HI boundary reaches the core surface and catches up with the shock already propagating in the core, resulting in a single stronger converging shock wave. 

The shock wave is reflected at the centre of the clump, and eventually gets back to the core edge. The contraction is almost halted, and since the core has a much higher density than the surrounding medium, it starts to expand. The expansion velocity $v_{exp}$ is computed considering the discontinuity between the core compressed by the reflected shock wave and the ICM at rest, using the arbitrary discontinuity algorithm.

We have explicitly verified that the core is so dense ($n\simeq 10^{5-6}$ cm$^{-3}$) during the contraction phase that the FUV radiation penetrates to a negligible depth with respect to its radius. Thus we can ignore photoevaporation during the contraction phase. When the clump starts expanding, we have computed for each time $t$ the thickness $\delta_\hi (t)$ of an HI shell (see eq. \ref{deltaHI}) for the corresponding core gas density. We get the core radius at $t$ by subtracting $\delta_\hi (t)$ to the radius $R(t) = R_0 + v_{exp}t$ ($R_0$ is the core radius at the end of the shock-contraction phase). 

\subsection{Stellar case} 

\begin{figure}
\includegraphics[scale=0.41]{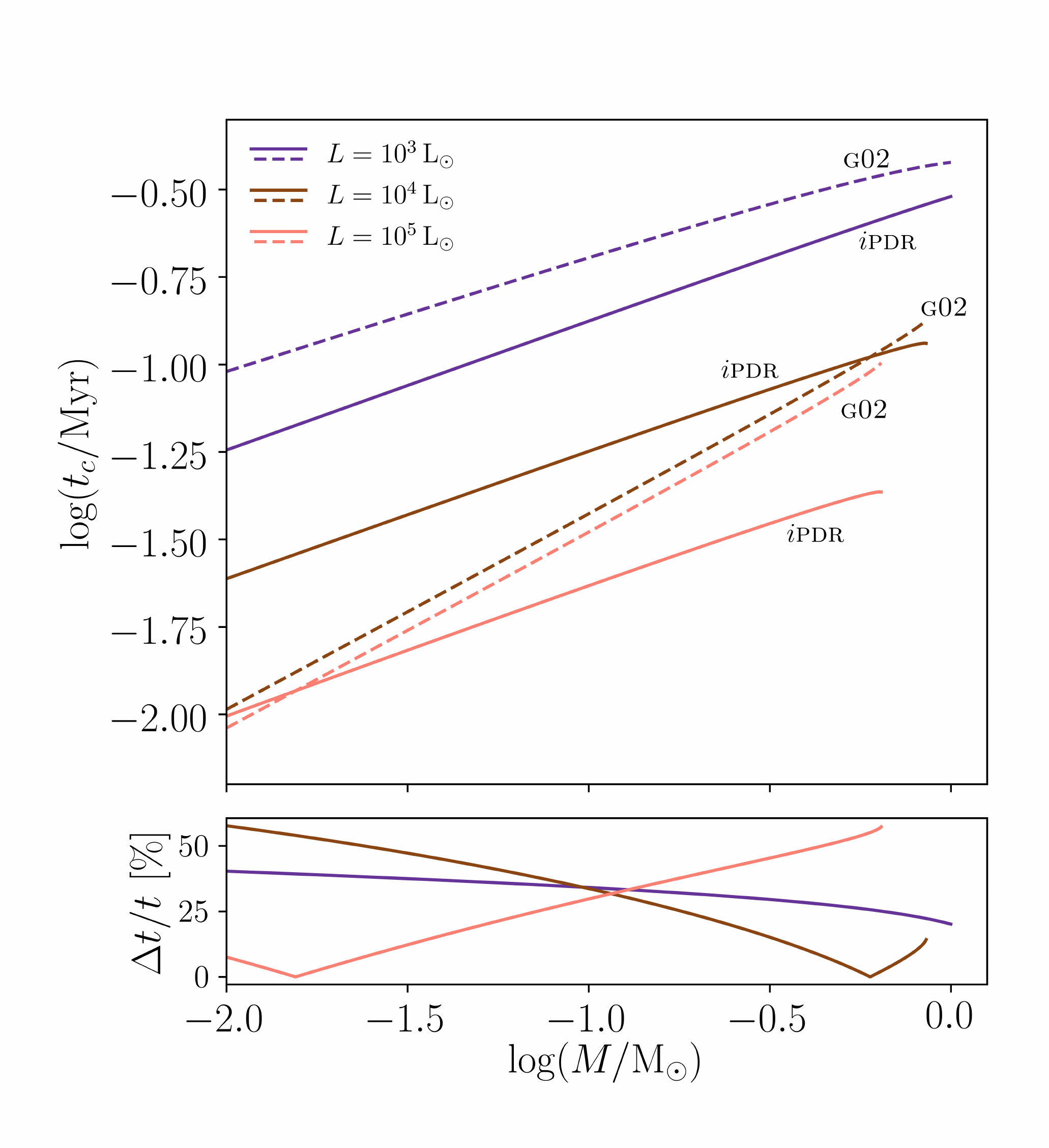}
\caption{{\bf Upper}: Clump evaporation time as a function of clump mass in the stellar case for different source luminosities, for fixed distance from the source ($0.3$ pc). The solid line representes our modified iPDR model, where we neglect ionizing radiation, the dotted line is the \citet{Gorti2002} model. {\bf Lower}: Relative difference between the two models $\Delta t/t = (t_\textsc{g02}-t_{i\textsc{pdr}})/\max(t_\textsc{g02}, t_{i\textsc{pdr}})$.}
\label{Fig5a}
\end{figure}

As a first application of our analysis, we consider molecular clumps photoevaporating because of stellar radiation. We consider a cold clump ($T_\h2 = 10$ K) located in the surrounding of a star, embedded in an atomic region with density $n_\textsc{at} = 10^3$ cm$^{-3}$. Then we assume that the expanding HII region of the massive star engulfs the clump (the density of the HII region gas is $n_\icm = 10\,\cc$, see Tab. \ref{tabIC}), and we apply the machinery we developed in Sec. \ref{Met}. We consider stars with bolometric luminosities $L=1\times 10^3\,\lsun$, $L=1\times 10^4\,\lsun$, $L=5\times 10^4\,\lsun$ and $L=1\times 10^5\,\lsun$. We assume the clump is located $0.3$ pc from the source, since this distance is smaller than the Str\"{o}mgren radius for every star in our set (for the fainter star the Str\"{o}mgren radius is $R_{Str}\simeq 0.75$ pc for a gas density $n=10\,\cc$). The BE masses for the collapse of molecular clumps at 10 K are around few tenths of solar masses, and for each luminosity we consider only clumps with mass below that limit.

The time evolution of the molecular core radius is shown in Fig. \ref{Fig5} (left panel), where a clump of initial mass $M=0.1\,\msun$ is exposed to the stellar radiation field for the different luminosities considered. The radius has a similar evolution for the different luminosities, with a shorter time-scales for larger luminosities. Consider for example the (brown) curve for $L=10^4\,\lsun$. A clump with mass $0.1\,\msun$ at the distance of $0.3$ pc has an initial radius of $0.023$ pc when it is in pressure equilibrium with the ICM. In the shock contraction phase, the radius reduces to $6\times 10^{-4}$ pc in about $6000$ yr because of the shock waves driven by the heated HI and HII shells. Then the expansion phase follows, and the core expands allowing the impinging radiation to penetrate and dissociate the molecules. This occurs significantly after the radius reaches its maximum value $r\simeq 0.025$ pc. 

While the contraction phase has almost the same duration for the tracks of the three more intense sources, we see that it takes more time for the $10^3\,\lsun$ star. In fact for this source the ionization fraction is low (see Fig. \ref{Fig1b}), since the temperature of the HII shell is only about 900 K. The shock driven by the HII shell is weak for this star, and needs more time to reach the centre of the core.

The lifetime of a clump ($t_c$) is defined as the time when the core radius goes to zero. In Fig. \ref{Fig5} (right panel) we show $t_c$ as a function of the clump mass and the source luminosity, at the same distance to the source ($0.3$ pc). 

\citet[][hereafter \G02cit]{Gorti2002} compute the lifetime of clumps located in a stellar PDR, in the absence of ionizing radiation. They account for photoevaporation by assuming that the clump continuously loses mass at a rate
\begin{equation}
	\dfrac{{\rm d} M}{{\rm d} t} = - 4\pi\rho_c r^2_c(t)c_\pdr\,\,,
\end{equation}
where $\rho_c$ is the mean mass density of the clump, and $c_\pdr$ the sound speed in the PDR of the clump. This implies that the clump loses mass also in the shock-compression phase. \G02cit do not account for the shock reflection at the centre of the clump, so that the core does not expand after the compression phase. On the other hand, in our treatment, photoevaporation is negligible while the clump is being compressed, and the shock reflection allows for the following expansion of the core. As a result, radiation is allowed to penetrate and dissociate the molecules only when the gas is sufficiently expanded and diluted. Furthermore, \G02cit find 
that under certain initial conditions\footnote{According to \citet{Gorti2002}, a clump undergoes a shock-compression only if its mean column density is \[nr_c < 2.7\times  10^{21}\,{\rm cm}^{-2}\,(c_\pdr/c_\h2)^3\,\,.\]} there is no shock-compression, since the shock suddenly stalls just after its formation, and the clump directly expands and photoevaporates. In our treatment, we do not recover this scenario, since we always allow the shock to reach the centre of the core. Magnetic and turbulent contribution to pressure are included by \G02cit, but not in this work. 

In Fig. \ref{Fig5a} we compare our predictions for the photoevaporation time (without ionizing radiation) with those from the \G02cit model. A range of clump masses between $0.01 \,\msun$ and the BE mass is considered, at a distance of $0.3$ pc, for three different source luminosities, A modification in our code for iPDR is required, since the HII shell is not present when ionizing radiation is absent. In the \G02cit model, the $10^4\,\lsun$ and the $10^5\,\lsun$ sources induce a shock-compression in the clump, while instead they predict an initial expansion for the $10^3\,\lsun$ case. The evaporation time-scale for the low-$L$ case differs by an order of magnitude with respect to the other two. In our model we do not find such dichotomy, and the evaporation time smoothly increases with $L$. However, the lifetimes are in agreement within a factor of 2 with those found in \G02cit.

Finally, we make a comparison between the evaporation times obtained with our full iPDR model (see Fig. \ref{Fig5}) and our model without ionizing radiation, i.e. with no HII shell (see Fig. \ref{Fig5a}). Clump lifetimes are always shorter when we consider the ionizing part of the spectrum, generally by a factor between 2 and 4 depending on clump mass and luminosity. This behaviour is expected, since the outer shell of the clump is heated to an high temperature and a stronger and faster shock propagates into the clump, decreasing its evolution time-scale.

\subsection{Quasar case} 

\begin{figure*}
\centering
\includegraphics[width=0.49\textwidth]{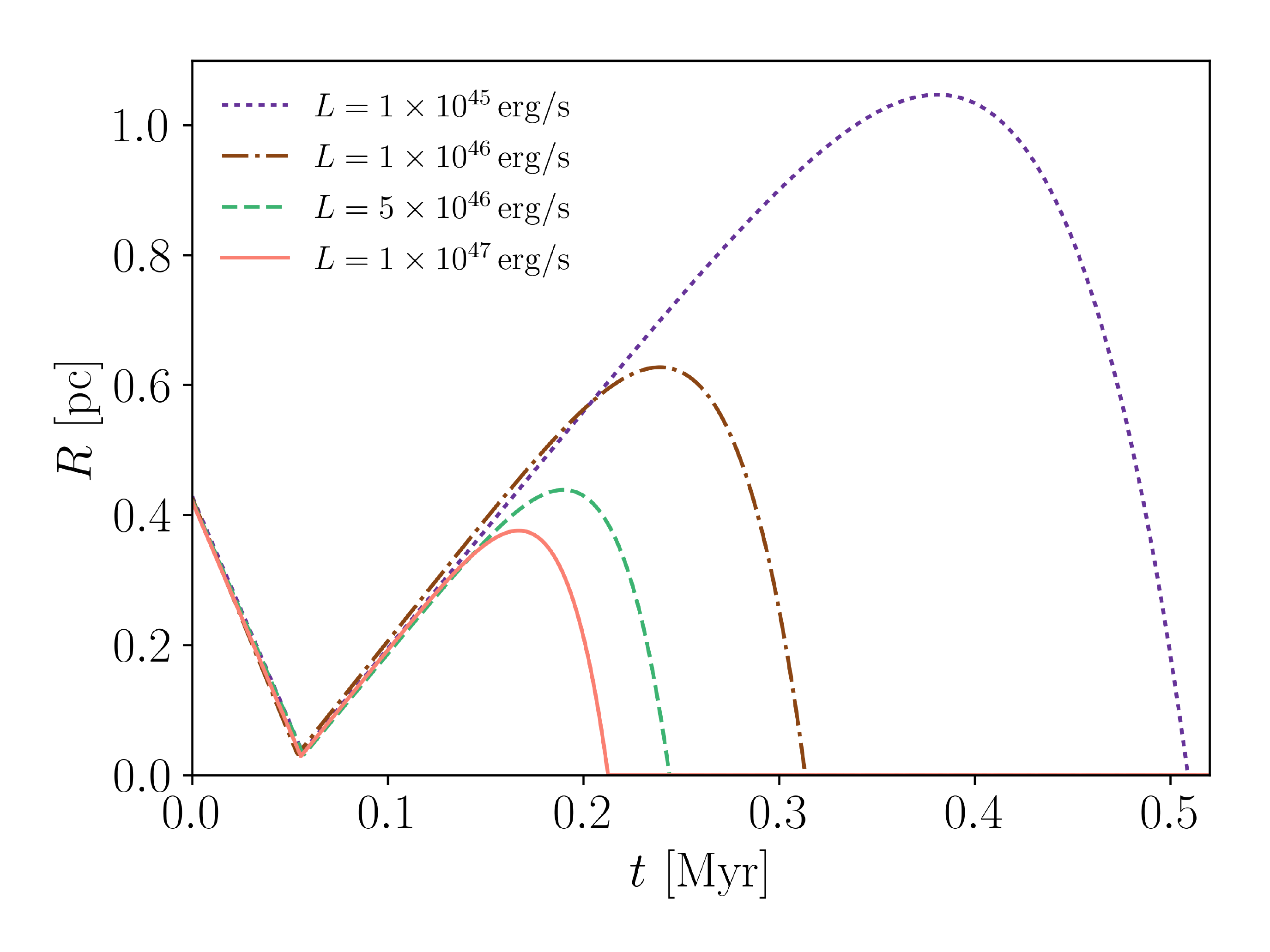}
\includegraphics[width=0.5\textwidth]{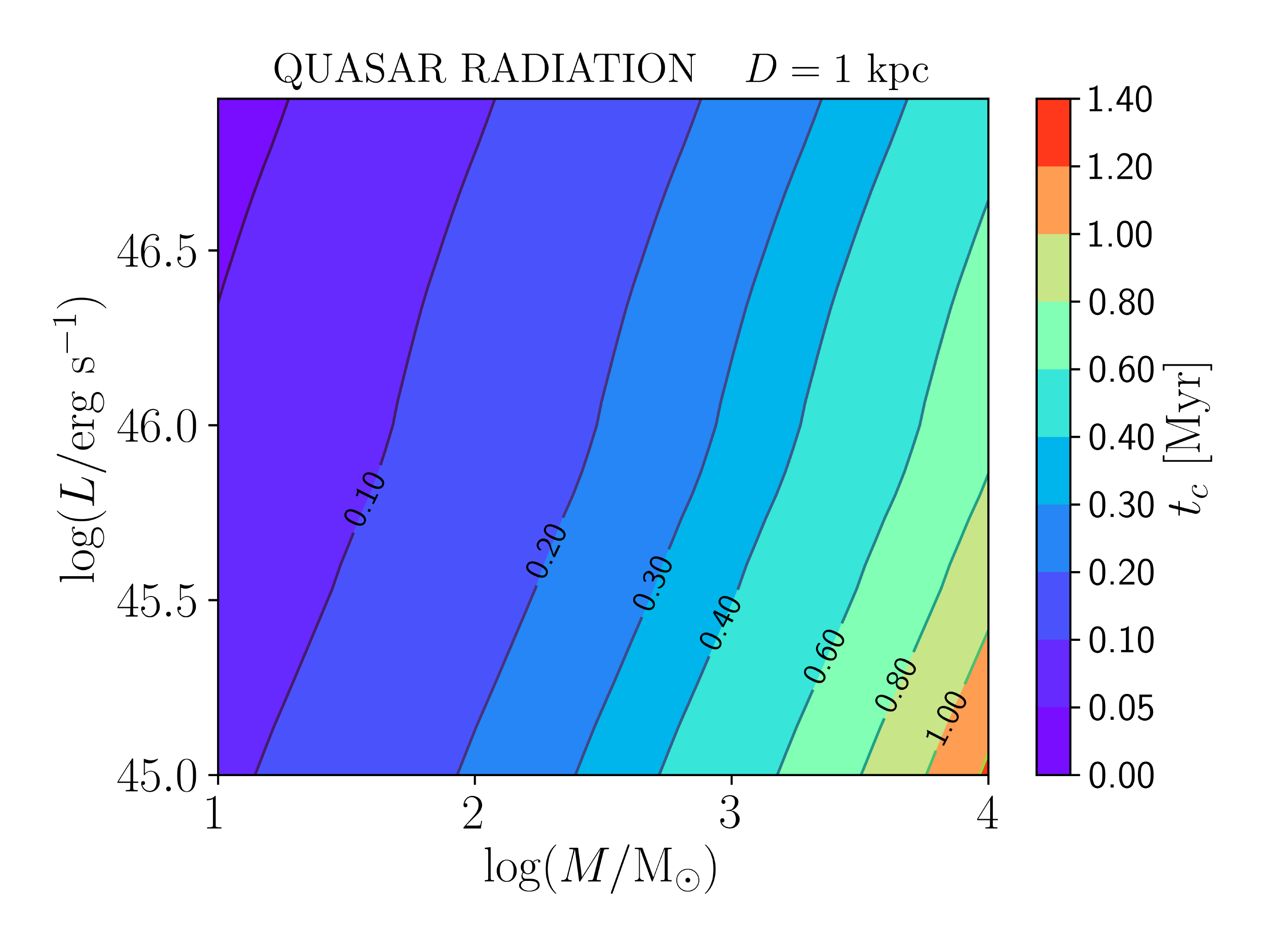}
\caption{{\bf Left} panel: radius ($R$) of the molecular core of a clump as a function of time, when the clump is exposed to quasar radiation. The clump has a mass of $10^3\, \msun$ and different luminosities are considered. {\bf Right} panel: Lifetime ($t_c$) of a clump exposed to the radiation of a quasar, for a range of values of the initial clump mass and different source luminosities. The distance of the clump from the source is fixed at 1 kpc.}
\label{Fig7}
\end{figure*}

\begin{table}
\centering
\begin{tabular}{cccc}
\hline \hline
\multicolumn{1}{c|} \, & $L$ & $F_{ion}\,^\star$ & $G_0$ \\ \hline
\multirow{3}{*}{\shortstack{OB star \\ {$(0.3\, {\rm pc})$}}} & \multicolumn{1}{c|} {$10^3\, \lsun$} & $0.006\,$ & $80$ \\
					  & \multicolumn{1}{c|} {$10^4\, \lsun$} & $0.5$ & $1.2\times 10^3$ \\
					  & \multicolumn{1}{c|} {$10^5\, \lsun$} & $14$ & $9.5\times 10^3$ \\\hline
\multirow{3}{*}{\shortstack{Quasar \\ {$(1\, {\rm kpc})$}}} & \multicolumn{1}{c|} {$10^{45}\, {\rm erg}\,{\rm s}^{-1}$} & $3.4$ & $6\times 10^2$ \\
					 & \multicolumn{1}{c|} {$10^{46}\, {\rm erg}\,{\rm s}^{-1}$} & $34$ & $6\times 10^3$ \\
					 & \multicolumn{1}{c|} {$10^{47}\, {\rm erg}\,{\rm s}^{-1}$} & $340$ & $6\times 10^4$ \\ \hline 
\end{tabular}
\caption{FUV and ionizing fluxes at a distance $D=0.3$ pc in stellar and $D=1$ kpc in the quasar case. $F_{ion}$ is in units of ${\rm erg}\,{\rm s}^{-1}\,{\rm cm}^{-2}$, $G_0$ is in units of the Habing flux.}
\label{fluxes}
\end{table}

\begin{figure}
\includegraphics[scale=0.31]{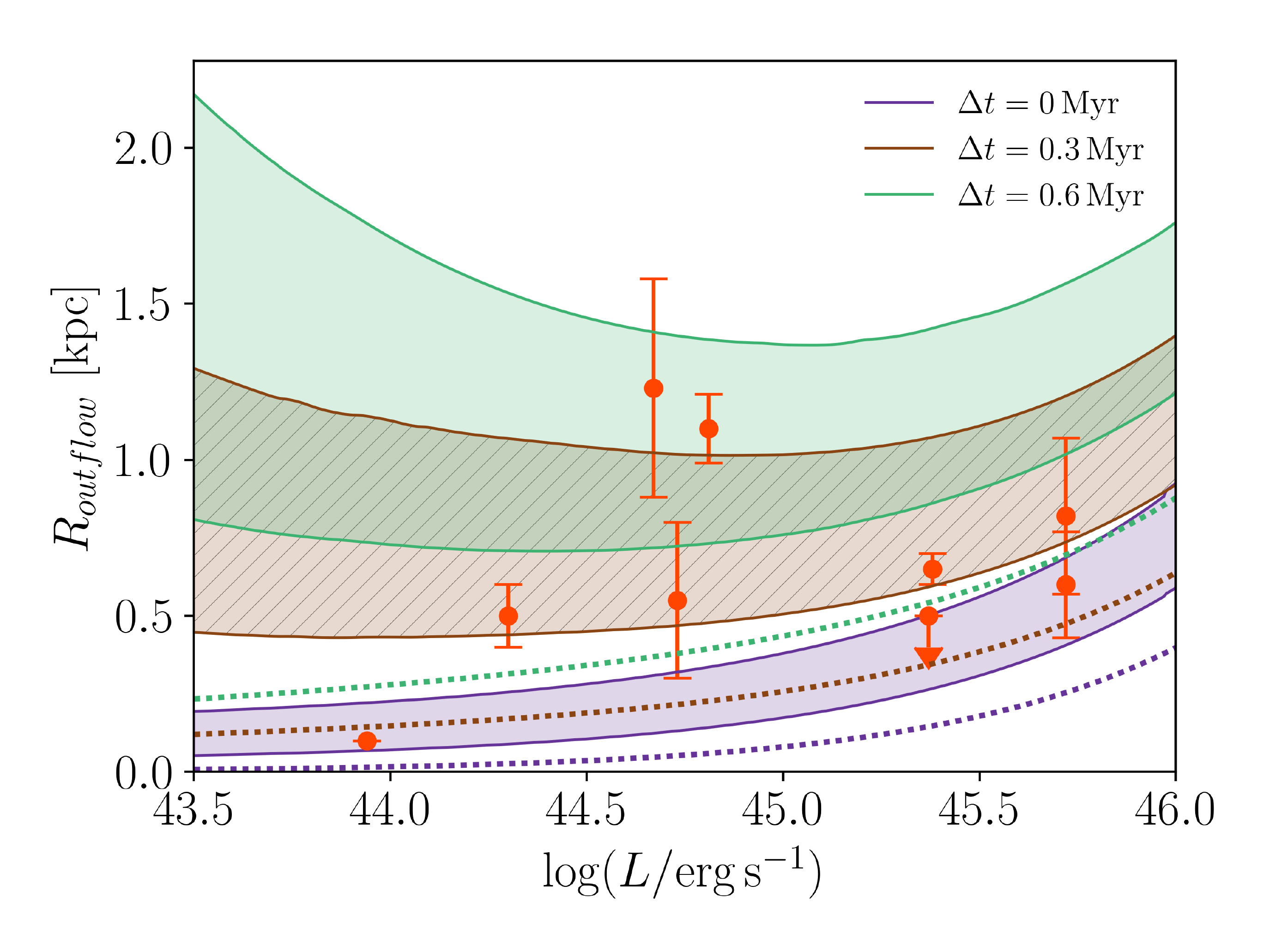}
\caption{ Extension of the molecular outflow for quasars with a range of luminosities, according to our photoevaporation model, assuming that clumps with masses in the range $M_c=0.1 - 0.9\,\,M_\textsc{be}$ ($M_\textsc{be}$ is the BE mass for collapse) form at the contact discontinuity (CD) between the wind and the ISM around the quasar. The violet dotted line is the initial position of the CD (i.e. the critical radius of the outflow), while the brown dotted and the green dotted lines are the position of the CD after a time $\Delta t = 0.3$ Myr and $\Delta t = 0.6$ Myr respectively. Shaded regions are the maximum distance that a clumps with masses in the considered range can travel before being completely photoevaporated, as a function of quasar luminosity, assuming that they form at the corresponding (same line color) CD position.}
\label{Fig8}
\end{figure}

We now describe the evolution of a clumps forming in the ionized outflows of quasars. We choose 1 kpc as a typical distance of a molecular clump from the source, that is of the order of the critical radius (eq. \ref{criticalradius}). The mass of clumps has been estimated by \citet{Zubovas2014a} to be around $8600 \, \msun$, thus we consider masses up to $10^4 \, \msun$. The ICM is the hot ionized medium of the wind, with temperature $T_\icm ¸\simeq 2.2\times 10^7$ K, and density $n_\icm \simeq 60$ cm$^{-3}$ at 1 kpc from the source. As explained in Sec. \ref{icm}, we assume that the clump is in pressure equilibrium with the ICM until its temperature reaches $10^4$ K (when the clump gas is still in atomic form). Afterwards, the gas turns into molecular form ($T_\h2=100$ K) in a very short time-scale, so that the density profile remains unchanged with respect to the $10^4$ K BE sphere. We apply our model to quasars with bolometric luminosities $L=1\times 10^{45}\, {\rm erg}\,{\rm s}^{-1}$, $L=1\times 10^{46}\, {\rm erg}\,{\rm s}^{-1}$, $L=5\times 10^{46}\, {\rm erg}\,{\rm s}^{-1}$ and $L=1\times 10^{47}\, {\rm erg}\,{\rm s}^{-1}$, with a spectral energy distribution given by eq. \ref{quasarSED}.

The evolution of the molecular core radius of the clump is shown in the left panel of Fig. \ref{Fig7}, while the right panel shows the lifetime of a clump at a distance of 1 kpc from a quasar, as a function of clump mass and source luminosity. Similarly to clumps around stars, the core radius presents a contraction phase followed by an expansion phase, where the core is dissociated and then ionized. Notice that the evaporation time is about ten times longer than in the stellar case, although the clump mass considered is about $10^4$ times larger. This is because both the FUV and ionizing radiation field are much more intense since the quasar spectrum  extends to very high energies (see Table \ref{fluxes} for reference values of the fluxes). This implies a higher temperature of the HII and HI shells, and a stronger compression ratio of the shock waves originating at the discontinuities.

Observations have detected molecular gas only up to a maximum distance of few kpc from quasars \citep[][hereafter \C14cit]{Cicone2014}. Photoevaporation has sometimes been invoked as an explanation for such limited extension. We have slightly modified our code to  account for the fact that radiation intensity decreases as the clump moves away from the source, being carried by the outflow. $\mbox{\citet{Ferrara2016c}}$ have shown that molecular clumps forming at the base of the adiabatic outflow are ablated in a short time because of the friction by hot flowing gas. Therefore, we analyse the alternative scenario in which clumps form within the outflow, so that they are at rest with the outflow and they are not subject to a strong acceleration. In this way our model is able to predict the distance travelled by a clump during its lifetime, and we can compare this length with observations of molecular outflow extensions. 

We consider a subset of active galactic nuclei (AGNs) listed in \C14cit (Table \ref{QSOlist}). In Fig. \ref{Fig8} we plot the outflow extension obtained with our photoevaporation model as a function of source luminosity. We study clumps with 90\% of their BE mass for collapse (i.e. the most massive clumps that do not collapse), forming at the contact discontinuity (CD) between the quasar wind and the surrounding ISM. According to \citet{King2010} model, the initial position of the CD coincides with the critical radius $R_c$ given in eq. \ref{criticalradius}, and it moves at a speed
\begin{equation}
	v_\textsc{cd} = 875 \, \sigma_{200}^{2/3} \,{\rm km\,s^{-1}} 
\end{equation}
in the energy-driven phase. In Fig. \ref{Fig8} the dotted lines correspond to the position of the CD at different times, while the shaded regions represent the maximum distance that clumps with a range of masses (0.1-0.9 times their Bonnor-Ebert mass) can travel before photoevaporating. The observed extension of the outflows in the considered quasar sample exceed the maximum distance travelled by clumps before they are photoevaporated, if they form at $R_c$. This implies that there is no mechanism more efficient than photoevaporation in destroying molecular clumps. On the other hand, the existence of outflows with large extensions (up to 1 kpc) suggests that clumps continue to form within the outflow, when the CD moves outwards from its initial position. This can bee see from the other two cases shown in Fig. \ref{Fig8} where we consider also 
clumps formed at a later times when the CD has moved to a radius $R_c +v_\textsc{cd} \Delta t$, with $\Delta t=0.3 ,0.6$ Myr. It appears that such delayed formation via thermal instabilities in the outflow can match the observed extensions. 


\section{Conclusions}
\label{Con}

\begin{table}
\centering
\begin{tabular}{cccc}
\hline \hline
\multicolumn{1}{c|} {Object} & $\log(L\, [{\rm erg}\,{\rm s}^{-1}])$ & $R_{\rm H_2}\,{\rm [kpc]}$ & Reference \\ \hline
\multicolumn{1}{c|} {NGC 1068} & $43.94$ & $0.10$ & a\\
\multicolumn{1}{c|} {IC 5063} & $44.30$ & $0.50 \pm 0.10$ & b \\
\multicolumn{1}{c|} {IRAS 23365+3604} & $44.67$ & $1.23 \pm 0.35$ & c \\  
\multicolumn{1}{c|} {Mrk 273} & $44.73$ & $0.55 \pm 0.55$ & c \\
\multicolumn{1}{c|} {IRAS F10565+2448} & $44.81$ & $1.10 \pm 0.11$ & c \\
\multicolumn{1}{c|} {I Zw 1} & $45.37$ & $<0.50$ & c \\
\multicolumn{1}{c|} {NGC 6240} & $45.38$ & $0.65 \pm 0.05$ & d \\
\multicolumn{1}{c|}	{IRAS F08572+3915} & $45.72$ & $0.82 \pm 0.17$ & c \\
\multicolumn{1}{c|}	{Mrk 231} & $45.72$ & $0.60 \pm 0.25$ & e \\ \hline 
\end{tabular}
\caption{Active galaxies showing molecular outflowing gas, table extracted from \citet{Cicone2014}. References for the molecular outflow measurements: a) \citet{Krips2011}, b) \citet{Morganti2015}, c) \citet{Cicone2014}, d) \citet{Feruglio2013b}, e) \citet{Cicone2012a}. For the first object, the error is not mentioned in the referenced work, in the case of I Zw 1 the radius is an upper limit.}
\label{QSOlist}
\end{table}

We have studied the evolution of molecular clumps exposed to radiation having both a far ultraviolet (FUV) and an ionizing component, determining the formation of an ionization/photodissociation region (iPDR) at the surface of clumps. The cases of a clump forming in the surroundings of an OB stars and a clump forming in the fast outflow of a quasar are studied separately. The clump is assumed to be an isothermal Bonnor-Ebert sphere with a mass lower than the critical mass for collapse. We assume a sudden heating scenario, inducing a shell structure in the clump, and then we analyse the evolution of its radius and density profile as a function of time, finally computing the clump lifetime (i.e. the time at which the molecular gas in the clump is completely dissociated). The clump evolution is solely determined by two parameters: its mass, $M$, and the bolometric luminosity $L$ of the source. 

We show that the pressure difference between adjacent layers causes the propagation of shock and rarefaction waves into the clump. The core shrinks until the shock wave hits the centre and reflects back, while the external layers expand and become eventually transparent to radiation. The dense core is thus surrounded by a diluted medium and it starts an expansion phase. As a result, the core density decreases and the radiation propagates in the interior, progressively evaporating the whole core. In this analysis we have not included gravity effects which could limit the expansion following the shock-contraction phase of clumps. Gravity may also play a role for the clumps that become gravitationally unstable during the contraction phase, possibly triggering star formation \citep{Bisbas2011, Walch2012}.

In the stellar case, we find that a higher luminosity speeds up considerably the shock-contraction phase: clumps of 0.1 $\msun$ at 0.3 pc from the source evaporate in 0.01 Myr for the brightest star considered ($10^5 \, \lsun$), while it takes 0.06 Myr in the case of the $10^3 \, \lsun$ star. Indeed, the radiation from the fainter star is not able to completely ionize the surface layer of the clump, resulting in a lower pressure of the HII shell and a weaker shock-induce contraction phase. 

Our model agrees within a factor of 2 with the \citet[][\G02cit]{Gorti2002} model, in the case of clumps embedded in the PDR of a massive star and in absence of ionizing radiation. The main difference between the two models is the evaporation channel. In \G02cit evaporation is due to a constant mass flow from the clump surface; in our model the clump evaporates as a consequence of the expansion and dilution driven by the reflected shock wave. We also notice that, in the absence of ionizing radiation,  evaporation times are always longer by a factor $2-4$ with respect to the full iPDR model including both FUV and ionizing radiation. Therefore, considering ionizing radiation is important, since the evolution history of clumps is significantly modified.

In the context of high-redshift galaxies, this is significant for far infrared (FIR) emission, as [CII]. Indeed, most of the [CII] emission from high-redshift galaxy seems to be due to molecular clumps \citep{Yue2015, Vallini2015, Pallottini2017} and because of the high radiation field observed in such galaxies \citep{Inoue2016, Carniani2017}, photoevaporation can play an important role. While \citet{Vallini2017} analyses the effect of clump photoevaporation using a time evolution based on \G02cit, we argue that shorter photoevaporation time-scales obtained with our iPDR model could further affect the detectability of  high-redshift galaxies. However, we underline that other effects are also important: the contrast with cosmic microwave background (CMB) attenuates the observed FIR emission for redshift $z\gtrsim 5$ \citep{Dacunha2013, Zhang2016}, which is relevant for low density gas \citep[$n<0.1 \cc$,][]{Vallini2015, Pallottini2015}, while CO destruction by cosmic rays may enhance [CI] and [CII] emission \citep{Bisbas2015}.  

The evolution in the quasar context is characterized by a similar behaviour. The duration of the contraction phase is roughly constant for different $L$, since all the quasars in the set are able to completely ionize and heat to about $10^4$ K the outer shell of the clump. We obtain evaporation times of 0.21 Myr for the $10^{47}\,{\rm erg}\,{\rm s}^{-1}$ quasar and 0.51 Myr for the $10^{45}\,{\rm erg}\,{\rm s}^{-1}$ quasar. With comparison to the stellar case, the evaporation times are longer only by a factor $\sim 10$, even though the clumps in the quasar case are $\simeq 10^4$ times more massive. This is consistent with the higher UV fluxes produced by quasars in spite of the larger spatial scales of the problem (see Tab. \ref{fluxes}). 

Applying our algorithm to clumps embedded in quasar outflows, we have been able to predict the outflow extension. This is set by the maximum distance travelled by clumps before photoevaporating, assuming that they form at the contact discontinuity (CD) between the quasar wind and the ISM. We find that the observed molecular outflow extensions are always larger than the distance travelled by clumps forming at the initial position of CD, but they are compatible with clumps forming at the CD with a time delay $\Delta t\simeq 0 - 0.6$ Myr after the outflow has entered the energy-driven phase. Therefore, we argue that:
\begin{itemize}
\item photoevaporation must be a crucial mechanism involved in the evolution of molecular gas structures in quasars, since none of the observed outflows has a smaller extension than what predicted with our photoevaporation model;
\item clumps need to form continuously within outflows, when the CD has moved farther from the quasar, in order to explain the most extended outflows.
\end{itemize}
A more comprehensive analysis of quasar outflows should consider a distribution of clump masses, the contribution of scattered light in a clumpy medium and the possible occurrence of star formation within the outflow \citep{Maiolino2017}.

\bibliographystyle{style/mnras}
\bibliography{biblio}

\appendix

\section{Collision of rarefaction waves}
\label{RWcollision}

\begin{figure}
\centering
\includegraphics[width=0.47\textwidth]{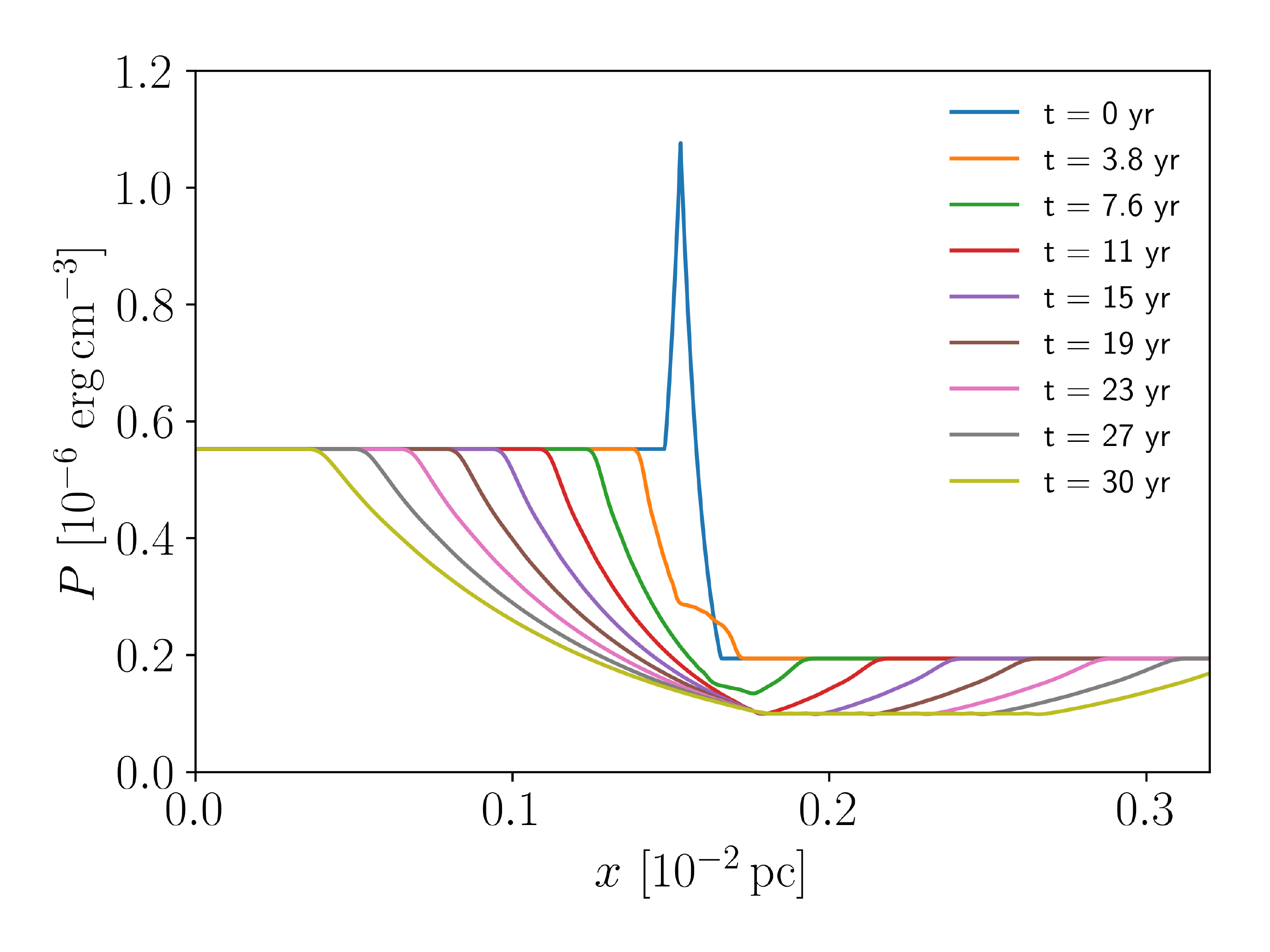}
\caption{Pressure profile of a gas at different timesteps, after the head-on collision of two rarefaction waves. The initial condition are chosen as for the collision happening in the HII shell of a $10^3\,\msun$ clump exposed to the radiation of a $10^{46}\, {\rm erg}\,{\rm s}^{-1}$ quasar (see Fig. \ref{Fig3}).}
\label{Fig9}
\end{figure}

The pressure and density profile of a fluid crossed by a rarefaction wave is given by eq. \ref{rarefactionprofile} (assuming an isothermal  process). Since the profile is not a discontinuity, as for shock waves, collisions between rarefaction waves cannot be studied analytically. The MacCormack method \citep{MacCormack2003} is a discretization algorithm for solving numerically hyperbolic differential. We apply this method to find a solution of fluid dynamics equation, where the initial condition is set by two approaching rarefaction wave, travelling in the opposite direction. In Fig. \ref{Fig9} we have simulated the collision of the two rarefaction waves generating at the edges of the HII shell of a clump exposed to the UV radiation of a quasar (see Fig. \ref{Fig3} and description in Sec. \ref{Res}). The two rarefaction waves are set so that they travel $3\times 10^{-4}$ pc before colliding, a distance of the order of the HII shell thickness (see Fig. \ref{Fig11}). The lines in figure represent the pressure and density profile at different intervals of time. The total length considered ($3\times 10^{-2}$ pc) has been divided in a $10^5$-cell grid, while the timesteps are chosen according to the Courant condition \citep{Courant1928}. The result of the collision is two rarefaction waves travelling in the opposite direction, with an equilibrium pressure in the central region 
\[P_\textsc{mc}\simeq 1.0\times 10^{-7}\, {\rm erg}\,{\rm cm}^{-3}\,\,.\] 
The same prediction for the outcome can be obtained with the arbitrary discontinuity algorithm discussed in Sec. \ref{shockdynamics}. We approximate the rarefaction waves to jumps in the flow variables ($P$, $\rho$, $v$), i.e. we consider a contact discontinuity between the post-rarefaction states of the fluid. We obtain for the equilibrium pressure a value 
\[P_\textsc{ad}\simeq 9.8\times 10^{-8}\, {\rm erg}\,{\rm cm}^{-3}\,\,,\]
in agreement with the value found previously within a few percent. Therefore, the discontinuity approximation introduces a small error, but it is more convenient from a computational point of view.

\label{lastpage}
\end{document}